\documentclass[12pt]{article}
\usepackage{a4wide}
\usepackage{amssymb}
\usepackage{amsbsy}
\usepackage{latexsym}
\usepackage{hyperref}
\usepackage{epsfig, psfrag, graphicx}

\newcommand{\be}{\begin{equation}}
\newcommand{\ee}{\end{equation}}
\newcommand{\ben}{\begin{eqnarray*}}
\newcommand{\een}{\end{eqnarray*}}
\newcommand{\bea}{\begin{eqnarray}}
\newcommand{\eea}{\end{eqnarray}}
\newcommand{\bdm}{\begin{displaymath}}
\newcommand{\edm}{\end{displaymath}}
\newcommand{\ba}{\begin{align}}
\newcommand{\ea}{\end{align}}
\newcommand{\lb}{\label}

\begin{document}
{\renewcommand{\thefootnote}{\fnsymbol{footnote}}
\hfill  IGC--08/4--2\\
\medskip
\begin{center}
{\LARGE  Effective Constraints for Quantum Systems}\\
\vspace{1.5em}
Martin Bojowald,\footnote{e-mail address: {\tt bojowald@gravity.psu.edu}}$^1$
Barbara Sandh\"ofer,\footnote{e-mail address: {\tt bs324@thp.Uni-Koeln.DE}}$^{2,1}$
Aureliano Skirzewski\footnote{e-mail address: {\tt askirz@gmail.com}}$^3$
and Artur Tsobanjan\footnote{e-mail address: {\tt axt236@psu.edu}}$^1$
\\
\vspace{0.5em}
$^1$Institute for Gravitation and the Cosmos,
The Pennsylvania State
University,\\
104 Davey Lab, University Park, PA 16802, USA\\
\vspace{0.5em}
$^2$Institute for Theoretical Physics, University of
Cologne,\\Z\"ulpicher Strasse 77, 50937 Cologne, Germany
\\
\vspace{0.5em}
$^3$
Centro de F\'isica Fundamental,
Universidad de los Andes, M\'erida 5101, Venezuela
\vspace{1.5em}

\end{center}
}

\setcounter{footnote}{0}

\newtheorem{theo}{Theorem}
\newtheorem{lemma}{Lemma}
\newtheorem{defi}{Definition}

\newcommand{\proofend}{\raisebox{1.3mm}{\fbox{\begin{minipage}[b][0cm][b]{0cm}
\end{minipage}}}}
\newenvironment{proof}{\noindent{\it Proof:} }{\mbox{}\hfill \proofend\\\mbox{}}
\newenvironment{ex}{\noindent{\it Example:} }{\medskip}
\newenvironment{rem}{\noindent{\it Remark:} }{\medskip}

\newcommand{\case}[2]{{\textstyle \frac{#1}{#2}}}
\newcommand{\lP}{\ell_{\mathrm P}}

\newcommand{\md}{{\mathrm{d}}}
\newcommand{\tr}{\mathop{\mathrm{tr}}}
\newcommand{\sgn}{\mathop{\mathrm{sgn}}}
\newcommand{\del}{\partial}
\newcommand{\der}{^{\prime}}
\newcommand{\const}{\mathrm{const.}}
\newcommand{\Mmkl}{\sum_{m=0}^{n}\sum_{k=0}^{m}\sum_{\ell=0}^{n-m}{n
    \choose m}{m \choose k}{n-m \choose\ell}(-1)^{n-m}}
\newcommand{\Mm}{\sum_{m=0}^{n}{n\choose m}(-1)^{n-m}}
\newcommand{\Mmjl}{\sum_{m=0}^{n}\sum_{j=0}^{m}\sum_{\ell=0}^{2(n-m)}{n
    \choose m}{m \choose j}{2(n-m) \choose\ell}p_t^{m-j}\frac{p^{2(n-m)-\ell}}{(2M)^{n-m}}}
\newcommand{\Mmjlr}{\sum_{m=0}^{n}\sum_{j=0}^{m}\sum_{\ell=0}^{2(n-m)}\sum_{r=0}^{k}{n
    \choose m}{m \choose j}{2(n-m)
    \choose\ell}{k \choose r}p_t^{m-j}\frac{p^{2(n-m)+k-\ell-r}}{(2M)^{n-m}}}

\newcommand*{\R}{{\mathbb R}}
\newcommand*{\N}{{\mathbb N}}
\newcommand*{\Z}{{\mathbb Z}}
\newcommand*{\Q}{{\mathbb Q}}
\newcommand*{\C}{{\mathbb C}}

\begin{abstract}
 An effective formalism for quantum constrained systems is presented
 which allows manageable derivations of solutions and observables,
 including a treatment of physical reality conditions without
 requiring full knowledge of the physical inner product. Instead of a
 state equation from a constraint operator, an infinite system of
 constraint functions on the quantum phase space of expectation values
 and moments of states is used. The examples of linear constraints as
 well as the free non-relativistic particle in parameterized form
 illustrate how standard problems of constrained systems can be dealt
 with in this framework.
\end{abstract}

\section{Introduction}

Effective equations are a trusted tool to sidestep some of the
mathematical and conceptual difficulties of quantum theories. Quantum
corrections to classical equations of motion are usually easier to
analyze than the behavior of outright quantum states, and they can
often be derived in a manageable way. This is illustrated, e.g., by
the derivation of the low-energy effective action for anharmonic
oscillators in \cite{EffAcQM} or, equivalently, effective equations
for canonical quantum systems in \cite{EffAc,EffectiveEOM,Karpacz}. But
effective equations are not merely quantum corrected classical
equations. They provide direct solutions for quantum properties such
as expectations values or fluctuations. While semiclassical regimes
play important roles in providing useful approximation schemes,
effective equations present a much more general method. In fact, they
may be viewed as an analysis of quantum properties independently of
specific Hilbert space representation issues.

As we will discuss here, this is especially realized for constrained
systems which commonly have additional complications such as the
derivation of a physical inner product or the problem of time in
general relativity \cite{KucharTime}.  We therefore develop an
effective constraint formalism parallel to that of effective equations
for unconstrained systems. Its advantages are that (i) it avoids
directly writing an integral (or other) form of a physical inner
product, which is instead implemented by reality conditions for the
physical variables; (ii) it shows when a phase space variable evolves
classically enough to play the role of internal time which, in a
precise sense, emerges from quantum gravity; and (iii) it directly
provides physical quantities such as expectation values and
fluctuations as relational functions of internal time, rather than
computing a whole wave function first and then performing
integrations. These advantages avoid conceptual problems and some
technical difficulties in solving quantum equations. They can also
bring out general properties more clearly, especially in quantum
cosmology.  Moreover, they provide equations which are more easily
implemented numerically than equations for states followed by
integrations to compute expectation values.  (Finally, although we
discuss only systems with a single classical constraint in this paper,
anomaly issues can much more directly be analyzed at the effective
level; see \cite{Vector,Tensor,ScalarGaugeInv} for work in this
direction.)

As we will see, however, there are still various unresolved
mathematical issues for a completely general formulation. In this
article, we propose the general principles behind an effective
formulation of constrained systems and illustrate properties and
difficulties by simple examples, including the parameterized free,
non-relativistic particle where we will demonstrate the interplay of
classical and quantum variables as it occurs in constrained systems.
Specific procedures used in this concrete example will be general
enough to encompass any non-relativistic system in parameterized
form. Relativistic systems show further subtleties and will be dealt
with in a forthcoming paper.

\section{Setting}
\label{S:Setting}

We first review the setup of effective equations for unconstrained
Hamiltonian systems \cite{EffAc,EffectiveEOM,Karpacz}, which we will
generalize to systems with constraints in the following section.

We describe a state by its moments rather than a wave function in a
certain Hilbert space representation. This has the immediate advantage
that the description is manifestly representation independent and
deals directly with quantities of physical interest, such as
expectation values and fluctuations. Just as a Hilbert space
representation, the system is determined through the algebra of its
basic operators and their $\star$-relations (adjointness or reality
conditions). In terms of expectation values, fluctuations and all
higher moments, this structure takes the form of an infinite
dimensional phase space whose Poisson relations are derived from the
basic commutation algebra. Dynamics is determined by a Hamiltonian on
this phase space. As a function of all the phase space variables it is
obtained by taking the expectation value of the Hamiltonian operator
in a general state and expressing the state dependence as a dependence
on all the moments. Thus, the Hamiltonian operator determines a
function on the infinite dimensional phase space which generates
Hamiltonian evolution.\footnote{This viewpoint in the present context
of effective equations goes back to \cite{Time}. While some underlying
constructions can be related to the geometrical formulation of quantum
mechanics developed in \cite{GeomQuantMech,ClassQuantMech,Schilling},
the geometrical formulation has so far not provided a rigorous
derivation of effective equations. Present methods in this context
remain incomplete due to a lack of treating quantum variables
properly, which take center stage in the methods of \cite{EffAc} and
those developed here. In some cases, it may be enough to place upper
bounds on additional correction terms from quantum variables, based on
semiclassicality assumptions. This may be done within the geometric
formulation to provide {\em semiclassical equations}
\cite{Josh,ClassFromQuant}, but it is insufficient for {\em effective
equations}.}

Specifically, for an ordinary quantum mechanical system with canonical
basic operators $\hat{q}$ and $\hat{p}$ satisfying
$[\hat{q},\hat{p}]=i\hbar$ we have a phase space coordinatized by the
expectation values $q:=\langle\hat{q}\rangle$ and
$p:=\langle\hat{p}\rangle$ as well as infinitely many quantum
variables\footnote{Notice that the notation used here differs from
that introduced in \cite{EffAc} because we found that the
considerations of the present article, in which several canonical
pairs are involved, can be presented more clearly in this way.}
\begin{equation} \label{QuantVars}
 G^{a,b}:=\left
 \langle(\hat{p}-\langle\hat{p}\rangle)^a(\hat{q}-\langle\hat{q}\rangle)^{b}
\right\rangle_{\rm Weyl}
\end{equation}
for integer $a$ and $b$ such that $a+b\geq2$, where the totally
symmetric ordering is used. For $a+b=2$, for instance, this provides
fluctuations $(\Delta q)^2=G^{0,2}=G^{qq}$ and $(\Delta
p)^2=G^{2,0}=G^{pp}$ as well as the covariance $G^{1,1}=G^{qp}$. As
indicated, for moments of lower orders it is often helpful to list the
variables appearing as operators directly. The symplectic structure is
determined through Poisson brackets which follow by the basic rule
$\{A,B\}= -i\hbar^{-1}\langle[\hat{A},\hat{B}]\rangle$ for any two
operators $\hat{A}$ and $\hat{B}$ which define phase space functions
$A:=\langle\hat{A}\rangle$ and $B:=\langle\hat{B}\rangle$. Moreover,
for products of expectation values in the quantum variables one simply
uses the Leibniz rule to reduce all brackets to the elementary ones.
General Poisson brackets between the quantum variables then satisfy
the formula\footnote{We thank Joseph Ochoa for bringing a mistake in
the corresponding formula of \cite{EffAc}, as well as its correction,
to our attention.}
\begin{eqnarray} \label{Gbrackets}
 \{G^{a,b},G^{c,d}\}\!\!\!\! &=&\!\!\!\!\!\! \sum_{r,s=0}^{\infty}
 (-{\textstyle\frac{1}{4}}\hbar^2)^{r+s} \sum_{j,k}
{a\choose j}{b\choose k}{c\choose k} {d\choose j}
G^{a+c-j-k,b+d-j-k}
 (\delta_{j,2r+1}\delta_{k,2s} -
\delta_{j,2r}\delta_{k,2s+1})\nonumber\\
&& -adG^{a-1,b}G^{c,d-1} +bcG^{a,b-1}G^{c-1,d}
\end{eqnarray}
where the summation of $j$ and $k$ is over the ranges $0\leq j\leq
\min(a,d)$ and $0\leq k\leq \min(b,c)$, respectively.  (For low order
moments, it is easier to use direct calculations of Poisson brackets
via expectation values of commutators.)  This defines the kinematics
of the quantum system formulated in terms of moments. The role of the
commutator algebra of basic operators is clearly seen in Poisson
brackets.

Dynamics is defined by a quantum Hamiltonian derived from the
Hamiltonian operator by taking expectation values. This results in a
function of expectation values and moments through the state used for
the expectation value. By Taylor expansion, we have
\begin{eqnarray}
 H_Q(q,p,G^{a,b}) &=&\langle H(\hat{q},\hat{p})_{\rm Weyl}\rangle=\langle
H(q+(\hat{q}-q),p+(\hat{p}-p))_{\rm Weyl}\rangle\nonumber\\
&=&H(q,p)+\sum_{a=0}^{\infty} \sum_{b=0}^{\infty}
\frac{1}{a!b!}
\frac{\partial^{a+b}
H(q,p)}{\partial p^a\partial q^b}G^{a,b} \label{HQ}
\end{eqnarray}
where we understand $G^{a,b}=0$ if $a+b<2$ and $H(q,p)$ is the
classical Hamiltonian evaluated in expectation values.  As written
explicitly, we assume the Hamiltonian to be Weyl ordered. If another
ordering is desired, it can be reduced to Weyl ordering by adding
re-ordering terms.

Having a Hamiltonian and Poisson relations of all the quantum
variables, one can compute Hamiltonian equations of motion
$\dot{q}=\{q,H_Q\}$, $\dot{p}=\{p,H_Q\}$ and
$\dot{G}^{a,b}=\{G^{a,b},H_Q\}$. This results in infinitely many
equations of motion which, in general, are all coupled to each
other. This set of infinitely many ordinary differential equations is
fully equivalent to the partial differential equation for a wave
function given by the Schr\"odinger equation. In general, one can
expect a partial differential equation to be solved more easily than
infinitely many coupled ordinary ones. Exceptions are solvable systems
such as the harmonic oscillator or the spatially flat quantum
cosmology of a free, massless scalar field \cite{BouncePert} where
equations of motion for expectation values and higher moments
decouple. This decoupling also allows a precise determination of
properties of dynamical coherent states \cite{BounceCohStates}. Such
solvable systems can then be used as the basis for a perturbation
theory to analyze more general systems, just like free quantum field
theory provides a solvable basis for interacting ones. In quantum
cosmology, this is developed in
\cite{BouncePot,QuantumBounce,Recollapse}. Moreover, semiclassical and
some other regimes allow one to decouple and truncate the equations
consistently, resulting in a finite set of ordinary differential
equations. This is easier to solve and, as we will discuss in detail
below, can be exploited to avoid conceptual problems especially in the
context of constrained systems.

\section{Effective constraints}

For a constrained system, the definition of phase space variables
(\ref{QuantVars}) in addition to expectation values of basic operators
is the same. For several basic variables, copies of independent
moments as well as cross-correlations between different canonical
pairs need to be taken into account. A useful notation, especially for
two canonical pairs $(q,p;q_1,p_1)$ as we will use it later, is
\begin{equation} \label{QuantVars2}
 G^{a,b}_{c,d}\equiv\langle(\hat p -p)^a(\hat q -q)^b (\hat p_1
 -p_1)^c(\hat q_1 -q_1)^d \rangle_{\rm Weyl}\,.
\end{equation}
Also here we will, for the sake of clarity, sometimes use a direct
listing of operators, as in $G^{qq}=G^{2,0}_{0,0}=(\Delta q)^2$ or the
covariance $G^q_{p_1}= G^{0,1}_{1,0}$, for low order moments.

We assume that we have a single constraint $\hat{C}$ in the quantum
system and no true Hamiltonian; cases of several constraints or
constrained systems with a true Hamiltonian can be analyzed
analogously.  We clearly must impose the {\em principal} quantum
constraint $C_Q(q,p,G^{a,b}):=\langle\hat{C}\rangle=0$ since any
physical state $|\psi\rangle$, whose expectation values and moments we
are computing, must be annihilated by our constraint,
$\hat{C}|\psi\rangle=0$. Just as the quantum Hamiltonian $H_Q$ before,
the quantum constraint can be written as a function of expectation
values and quantum variables by Taylor expansion as in (\ref{HQ}).
However, this one condition for the phase space variables is much
weaker than imposing a Dirac constraint on states,
$\hat{C}|\psi\rangle=0$. In fact, a simple counting of degrees of
freedom shows that additional constraints must be imposed: One
classical constraint such as $C=0$ removes a pair of canonical
variables by restricting to the constraint surface and factoring out
the flow generated by the constraint. For a quantum system, on the
other hand, we need to eliminate infinitely many variables such as a
canonical pair $(q,p)$ together with all the quantum variables it
defines. Imposing only $C_Q=0$ would remove a canonical pair but leave
all its quantum variables unrestricted. These additional variables are
to be removed by infinitely many further constraints.

There are obvious candidates for these constraints. If
$\hat{C}|\psi\rangle=0$ for any physical state, we do not just have
a single constraint $\langle\hat{C}\rangle=0$ but infinitely many {\em
  quantum constraints}
\begin{eqnarray}
 C^{(n)} &:=& \langle\hat{C}^n\rangle =0\\
 C_{f(q,p)}^{(n)} &:=& \langle
f(\hat{q},\hat{p})\hat{C}^n\rangle =0 \label{Cfn}
\end{eqnarray}
for positive integer $n$ and arbitrary phase space functions
$f(q,p)$. All these expectation values vanish for physical states, and
in general differ from each other on the quantum phase space. For
arbitrary $f(q,p)$, there is an uncountable number of constraints
which should be restricted suitably such that a closed system of
constraints results which provides a complete reduction of the quantum
phase space. The form of functions $f(q,p)$ to be included in the
quantum constraint system depends on the form of the classical
constraint and its basic algebra. Examples and a general construction
scheme are presented below.

We thus have indeed infinitely many constraints,\footnote{As observed
in \cite{GeomConstr}, a single constraint $C^{(2)}$ would guarantee a
complete reduction for a system where zero is in the discrete part of
the spectrum of a self-adjoint $\hat{C}$. In this case, non-degeneracy
of the inner product ensures that
$\langle\psi|\hat{C}^2|\psi\rangle=0$ implies
$\hat{C}|\psi\rangle=0$. However, details of the quantization and the
quantum representation are required for this conclusion, based also on
properties of the spectrum, which is against the spirit of effective
equations. Moreover, the resulting constraint equation $C^{(2)}$ is in
general rather complicated and must be approximated for explicit
analytical or numerical solutions. Then, if $C^{(2)}=0$ is no longer
imposed exactly, a large amount of freedom for uncontrolled deviations
from $\hat{C}|\psi\rangle=0$ would open up. In our approach, we are
using more than one constraint which ensures that even under
approximations the system remains sufficiently well
controlled. Moreover, our considerations remain valid for constraints
with zero in the continuous parts of their spectra, although as always
there are additional subtleties.}  which constitute the basis for our
effective constraints framework. This is to be solved as a classical
constrained system, but as an infinite one on an infinite dimensional
phase space. An effective treatment then requires approximations whose
explicit form depends on the specific constraints. At this point, some
caution is required: approximations typically entail disregarding
quantum variables beyond a certain order to make the system
finite. Doing so for an order of moments larger than two results in a
Poisson structure which is not symplectic because only the expectation
values form a symplectic submanifold of the full quantum phase space,
but no set of moments to a certain order does. We are then dealing
with constrained systems on Poisson manifolds such that the usual
countings of degrees of freedom do not apply. For instance, it is not
guaranteed that each constraint generates an independent flow even if
it weakly commutes with all other constraints which would usually make
it first class. Properties of constrained systems in the more general
setting of Poisson manifolds which need not be symplectic are
discussed, e.g., in \cite{brackets}.

We also emphasize that gauge flows generated by quantum constraints on
the quantum phase space play important roles, which one may not have
expected from the usual Dirac treatment of constraints. There, only a
constraint equation is written for states, but no gauge flow on the
Hilbert space needs to be factored out. In fact, the gauge flow which
one could define by $\exp(it\hat{C})|\psi\rangle$ for a self-adjoint
$\hat{C}$ trivializes on physical states which solve the constraint
equation $\hat{C}|\psi\rangle=0$. In the context of effective
constraints, there are two main reasons why the gauge flow does not
trivialize and becomes important for a complete removal of gauge
dependent variables: First, to define the gauge flow
$\exp(it\hat{C})|\psi\rangle$ and conclude that it trivializes on
physical states, one implicitly uses self-adjointness of $\hat{C}$ and
assumes that physical states are in the kinematical Hilbert space for
otherwise it would not be the original $\hat{C}$ that could be used in
the flow. These are specific properties of the kinematical
representation which we are not making use of in the effective
procedure used here, where reality and normalization conditions are
not imposed before the very end of finding properties of states in the
physical Hilbert space. The expectation values and moments we are
dealing with when imposing quantum constraints thus form a much wider
manifold than the Hilbert space setting would allow. Here, not only
constraint equations but also gauge flows on the constraint surface
are crucial. If representation properties are given which imply that
physical states are in the kinematical Hilbert space, we will indeed
see that the flow trivializes as the example in Sec.~\ref{s:circle}
shows. Secondly, the Dirac constraint $\hat{C}|\psi\rangle=0$
corresponds to infinitely many conditions, and only when all of them
are solved can the gauge-flow trivialize. An effective treatment, on
the other hand, shows its strength especially when one can reduce the
required set of equations to finitely many ones, which in our case
would imply only a partial solution of the Dirac constraint. On these
partial solutions, which for instance make sure that fluctuations
correspond to those of a state satisfying $\hat{C}|\psi\rangle=0$ even
though other moments do not need to come from such a state, the
gauge-flow does not trivialize.

We will illustrate such properties as well as solution schemes of
effective constraints in examples below. But there are also general
conclusions which can be drawn. As the main requirements, we have to
ensure the system of effective constraints to be consistent and
complete. Consistency means that the set of all constraints should be
first class if we start with a single classical constraint or a first
class set of several constraints.  As we will illustrate by examples,
this puts restrictions on the form of quantum constraints, related to
the ordering of operators used, beyond the basic requirement that they
be zero when computed in physical states.

To show that the constraints are complete, i.e.\ they remove all
expectation values and quantum variables associated with one canonical
pair, we will consider a constraint $\hat{C}=\hat{q}$ in
Sec.~\ref{S:Cq}. Since locally one can always choose a single
(irreducible) constraint to be a phase space variable, this will serve
as proof that local degrees of freedom are reduced correctly. (Still,
global issues may pose non-trivialities since entire gauge orbits must
be factored out when constraints are solved.)

\subsection{The form of quantum constraints}
\lb{IterationConstraints}

At first sight, our definition of quantum constraints may seem
problematic. Some of them in (\ref{Cfn}) are defined as expectation
values of non-symmetric operators, thus implying complex valued
constraint functions. (We specifically do not order symmetrically in
(\ref{Cfn}) because this would give rise to terms where some $\hat{q}$
or $\hat{p}$ appear to the right while others remain to the left. This
would not vanish for physical states and therefore not correspond to a
constraint.) This may appear problematic, but one should note that
this reality statement is dependent on the (kinematical) inner product
used before the constraints are imposed. This inner product in general
differs from the physical one if zero is in the continuous part of the
spectrum of the constraint and thus reality in the kinematical inner
product is not physically relevant. Moreover, in gravitational
theories it is common or even required to work with constraint
operators which are not self-adjoint \cite{Komar}, and thus complex
valued constraints have to be expected in general. For physical
statements, which are derived after the constraints have been
implemented, only the final reality conditions of the physical inner
product are relevant.\footnote{At least partially, the meaning of
reality conditions depends on specifics of the measurement
process. This may be further reason to keep an open mind toward
reality conditions especially in quantum gravity.}

As we will discuss in more detail later, this physical reality can be
implemented effectively: We solve the constraints on the quantum phase
space, and then impose the condition that the reduced quantum phase
space be real. We will see explicitly that complex-valued quantum
variables on the unconstrained phase space are helpful to ensure
consistency. In parallel to Hilbert space notation, we call quantum
variables (\ref{QuantVars}) on the original quantum phase space {\em
kinematical quantum variables}, and those on the reduced quantum phase
space {\em physical quantum variables}. Kinematical quantum variables
are allowed to take complex values because their reality would only
refer to the inner product used on the kinematical Hilbert space. For
physical quantum variables in the physical Hilbert space as usually
defined, on the other hand, reality conditions must be imposed.

\subsubsection{Closure of constrained system}
\label{s:Closure}

Still, it may seem obvious how to avoid the question of reality of the
constraints altogether by using quantum constraints defined as
$G^{C^nf(q,p)}=\langle\hat C^n\widehat{f(p,q)}\rangle_{\rm Weyl}$ such
as $G^{C^nq}$ and $G^{C^np}$ with the symmetric ordering used as in
(\ref{QuantVars}).  Here, the symmetric ordering contained in the
definition of quantum variables must leave $\hat{C}$ intact as a
possibly composite operator, i.e.\ we have for instance $G^{C,p}=
\frac{1}{2}\langle\hat{C}\hat{p} +\hat{p}\hat{C}\rangle-Cp$
independently of the functional form of $\hat{C}$ in terms of
$\hat{q}$ and $\hat{p}$. Otherwise it would not be guaranteed that the
expectation value vanishes on physical states. We could not include
variables with higher powers of $q$ and $p$, such as $G^{C^npp}$ as
constraints because there would be terms in the totally symmetric
ordering (such as $\hat{p}\hat{C}^n\hat{p}$) not annihilating a
physical state. But, e.g., $G^{\hat{C}\hat{p}^2}$ understood as
$\frac{1}{2}\langle\hat{C}\hat{p}^2 +\hat{p}^2\hat{C}\rangle-Cp^2$
would be allowed. The use of such symmetrically ordered variables
would imply real quantum constraints.

However, this procedure is not feasible: The constraints would not
form a closed set and not even be first class. We have, for instance,
\begin{eqnarray*}
 \{G^{C^n,f(q,p)},G^{C^m,g(q,p)}\} &=&
 \frac{1}{4i\hbar}\langle[\hat{C}^n\hat{f}+ \hat{f}\hat{C}^n,
 \hat{C}^m\hat{g}+\hat{g}\hat{C}^m]\rangle\\
&& -\frac{g}{2i\hbar}\langle[\hat{C}^n\hat{f}+
 \hat{f}\hat{C}^n,\hat{C}^m]\rangle
 -\frac{C^m}{2i\hbar}\langle[\hat{C}^n\hat{f}+
 \hat{f}\hat{C}^n,\hat{g}]\rangle\\
&& -\frac{f}{2i\hbar}\langle[\hat{C}^n,\hat{C}^m\hat{g}+
 \hat{g}\hat{C}^m]\rangle
 -\frac{C^n}{2i\hbar}\langle[\hat{f},\hat{C}^m\hat{g}+
 \hat{g}\hat{C}^m]\rangle\\
&& +\{C^nf,C^mg\}\,.
\end{eqnarray*}
The first commutator contains several terms which vanish when the
expectation value is taken in a physical state, but also the two
contributions $[\hat{C}^n,\hat{g}]\hat{C}^m\hat{f}$ and
$\hat{f}\hat{C}^m[\hat{C}^n,\hat{g}]$ whose expectation value in a
physical state vanishes only if $\hat{f}$ or $\hat{g}$ commute with
$\hat{C}$. This would require quantum observables to be known and used
in the quantum constraints, which in general would be too restrictive
and impractical.

By contrast, the quantum constraints defined above do form a first
class system: We have
\begin{equation} \label{QuantConstrComm}
 [\hat{f}\hat{C}^n,\hat{g}\hat{C}^m] = [\hat{f},\hat{g}]\hat{C}^{n+m}
 +\hat{f}[\hat{C}^n,\hat{g}]\hat{C}^m
 +\hat{g}[\hat{f},\hat{C}^m]\hat{C}^n
\end{equation}
whose expectation value in any physical state vanishes. Thus, using
these constraints implies that their quantum Poisson brackets vanish
on the constraint surface, providing a weakly commuting set:
\begin{equation}
 \{C_f^{(n)},C_g^{(m)}\} = \frac{1}{i\hbar}
 \langle[\hat{f}\hat{C}^n,\hat{g}\hat{C}^m]\rangle \approx 0\,.
\end{equation}
A further possibility of using Weyl-ordered constraints of a specific
form will be discussed briefly in Sec.~\ref{s:Generate}, but also this
appears less practical in concrete examples than using non-symmetrized
constraints.

Constraints thus result for all phase space functions $f(q,p)$, but
not all constraints in this uncountable set can be independent. For
practical purposes, one would like to keep the number of allowed
functions to a minimum while keeping the system complete. Then,
however, the set of quantum constraints is not guaranteed to be closed
for any restricted choice of phase space functions in their
definition. If $C_f^{(n)}$ and $C_g^{(m)}$ are quantum constraints,
closure requires the presence of $C_{[f,g]}^{(n)}$ (for $n\geq 2$),
$C_{f[C^m,g]}^{(n)}$ and $C_{g[C^m,f]}^{(n)}$ as additional
constraints according to (\ref{QuantConstrComm}). This allows the
specification of a construction procedure for a closed set of quantum
constraints. As we will see in examples later, for a system in
canonical variables $(q,p)$ it is necessary to include at least
$C_q^{(n)}$ and $C_p^{(m)}$ in the set of constraints for a complete
reduction. With $C_{[q,p]}^{(n)}=i\hbar C^{(n)}$, the first new
constraints resulting from a closed constraint algebra add nothing
new. However, in general the new constraints $C_{q[C^m,p]}^{(n)}$ and
$C_{p[C^m,q]}^{(n)}$ will be independent and have to be
included. Iteration of the procedure generates further constraints in
a process which may or may not stop after finitely many steps
depending on the form of the classical constraint.

Although many independent constraints have to be considered for a
complete system, most of them will involve quantum variables of a
high degree. To a given order in the moments it is thus sufficient to
consider only a finite number of constraints which can be determined
and analyzed systematically. Such truncations and approximations will
be discussed by examples in Secs.~\ref{s:Trunc} and \ref{s:Approx}.

\subsubsection{Number of effective constraints: linear constraint operator}
\label{S:Number}

For special classes of constraints one can draw further conclusions at
a more general level. In particular for a linear constraint, which
shows the local behavior of singly constrained systems, it is
sufficient to consider polynomial multiplying functions as we will
justify by counting degrees of freedom.  Because this counting depends
on the number of degrees of freedom, we generalize, in this section
only, our previous setting to a quantum system of $N+1$ canonical
pairs of operators $(\hat{q}^i,\hat{p}_i)_{i=1,\ldots,N+1}$ satisfying
the usual commutation relations $[\hat{q}^i,\hat{p}_j]=i \hbar
\delta^i_j$.  Furthermore, it is sufficient to consider only the case
where the constraint itself is one of the canonical variables. Given
any constraint operator $\hat{C}$, linear in the canonical variables,
we can always find linear combinations of the canonical operators
$\left((\hat{x}_i)_{i=1, \ldots, N}; \hat{q}, \hat{p}\right)$\ such
that $\hat{q}=\hat{C}$ and
\[
\left[\hat{q},\hat{p}\right] = i\hbar\quad,\quad
\left[\hat{q},\hat{x}_i\right] = \left[\hat{p}, \hat{x}_i\right]=0\quad,\quad
\left[\hat{x}_i, \hat{x}_j\right] = i\hbar \left(\delta_{i,j-N} -
\delta_{i-N,j}\right)
\]
i.e.\ $\hat{x}_i$ form an algebra of $N$ canonical pairs ($i=1,
\ldots, N$ and $i=N+1, \ldots, 2N$ corresponding to the
configuration and momentum operators, respectively).\footnote{The
linear combinations that would satisfy the above relations may be
obtained by performing a linear canonical transformation on the
operators (post-quantization). Such combinations are not unique, but
this fact is not important for the purpose of counting the degrees
of freedom.} For the rest of this subsection we assume the above
notation, so that our quantum system is parameterized by the
expectation values $q:=\langle \hat{q} \rangle$, $p:=\langle
\hat{p}\rangle$, $x_i:=\langle \hat{x}_i \rangle$, $i=1,\ldots,2N$\
and the quantum variables:
\begin{equation}
G^{a_1, a_2,\ldots, a_{2N}; b, c} :=
\left\langle(\hat{x}_1-x_1)^{a_1} \cdots
(\hat{x}_{2N}-x_{2N})^{a_{2N}}(\hat{p}-p)^b(\hat{q}-q)^c
 \right\rangle_{\rm Weyl} \label{eq:Qvariables}
\end{equation}
where the operator product is totally symmetrized.

As proposed, we include among the constraints all functions of the
form $C_f=\langle \hat{f} \hat{C} \rangle$, where $\hat{f}$ is now
any operator polynomial in the canonical variables. This proposition
is consistent with $\hat{C} |\psi \rangle = 0$ and the set of
operators of the form $\hat{f} \hat{C}$ is closed under taking
commutators. As a result the set of all such functions $C_f$ is
first-class with respect to the Poisson bracket induced by the
commutator. (${C}^{(n)}_f$\ is automatically included in the above
constraints through $C_{f'}$\ where $\hat{f}'=\hat{f}\hat{C}^{n-1}$,
which is polynomial in the canonical variables so long as $\hat{f}$\
is.)

In principle, we have an infinite number of constraints to restrict
an infinite number of quantum variables. To see how the degrees of
freedom are reduced, we proceed order by order. Variables of the order
$M$ in $N+1$ canonical pairs are defined as in
Eq.~(\ref{eq:Qvariables}), with $\sum_{i}a_i + b + c = M$. The total
number of different combinations of this form is the same as the
number of ways the positive powers adding up to $M$ can be distributed
between $2(N+1)$ terms, that is ${M+2(N+1)-1 \choose
2(N+1)-1}$. Solving a single constraint classically results in the
(local) removal of one canonical pair. Subsequent quantization of the
theory would result in quantum variables corresponding to $N$
canonical pairs. In the rest of the section we demonstrate that our
selected form of the constraints leaves unrestricted precisely the
quantum variables of the form $G^{a_1, \ldots, a_{2N}; 0, 0}$.

It is convenient to make another change in variables. We note that in
order to permute two non-commuting canonical operators in a product we
need to add $i\hbar$\ times a lower order product. Starting with a
completely symmetrized product of order $M$\ and iterating the
procedure we can express it in terms of a sum of unsymmetrized
products of orders $M$\ and below, in some pre-selected order. In
particular, we consider variables of the form:
\begin{equation}
F^{a_1, a_2,\ldots a_{2N}; b, c} :=\left\langle (\hat{x}_1)^{a_1} \cdots
(\hat{x}_{2N})^{a_{2N}}\hat{p}^b\hat{q}^c
\right\rangle\label{eq:Fvariables}
\end{equation}
It is easy to see that there is a one-to-one correspondence between
variables (\ref{eq:Qvariables}) (combined with the expectation
values) and (\ref{eq:Fvariables}), but the precise mapping is
tedious to derive and not necessary for counting. We can immediately
see that our constraints require $F^{a_1, a_2,\ldots a_{2N}; b, c}
\approx 0$\ for $c\neq 0$. Moreover, all of the constraints
$C_f=\langle \hat{f} \hat{C} \rangle$\ may be written as a combination
of the variables $F^{a_1, a_2,\ldots a_{2N}; b, c}$, $c\neq 0$\
(again, this can be seen by noting that we may rearrange the order of
operators in a product by adding terms proportional to lower order
products). There are still too many degrees of freedom left as none of
the variables $F^{a_1, a_2,\ldots a_{2N}; b, 0}$\ are constrained.

At this point, however, we are yet to account for the unphysical
degrees of freedom associated with the gauge transformations.
Indeed, every constraint induces a flow on the space of quantum
variables through the Poisson bracket, associated with the
commutator of the algebra of canonical operators. The set of
constraints $C_f$\ is first-class, which means that the flows they
produce preserve constraints and are therefore tangent to the
constraint surface. However, not all of the flow-generating vector
fields corresponding to the distinct constraints considered above
will be linearly independent on the constraint surface because, to
a fixed order in moments, we are dealing with a non-symplectic
Poisson manifold. The degeneracy becomes obvious when we count the
degrees of freedom to a given order. To order $M$\ the constraints
are accounted for by variables $F^{a_1, a_2,\ldots a_{2N}; b, c+1}$,
where $\sum_{i}a_i + b + c + 1 = M$. Counting as earlier in the
section, there are ${M+2(N+1)-2\choose 2(N+1)-1}$\ such variables.
Subtracting the number of constraints from the number of quantum
variables of order $M$, we are left with
\begin{eqnarray}
&&{M+2(N+1)-1 \choose 2(N+1)-1} - {M+2(N+1)-2 \choose 2(N+1)-1} \nonumber\\ 
&=&
\left(\frac{M+2(N+1)-1}{M+2(N+1)-1-(2N+1)} - 1 \right) {M+2(N+1)-2 
\choose 2(N+1)-1} \nonumber \\
&=& \frac{2(N+1)-1}{M}{M+2(N+1)-2 \choose 2(N+1)-1}
\label{eq:dof_wo_constr}
\end{eqnarray}
unrestricted quantum variables. If each constraint does generate an
independent non-vanishing flow, we should subtract the number of
constraints from the result again to get
$\frac{2(N+1)-1-M}{M}{M+2(N+1)-2 \choose 2(N+1)-1}$\ physical
degrees of freedom of order $M$. This number becomes negative once
$M$ is large enough raising the possibility that the system has
been over-constrained.

Fortunately, this is not the case. All of the operators $\hat{x}_i$
commute with the original constraint operator $\hat{C}$($\equiv \hat{q}$),
which means that any function of the expectation value of a polynomial
in $(\hat{x}_i)_{i=1,\ldots, 2N;}$\ $g = \langle g[\hat{x}_i] \rangle$, weakly
commutes with every constraint
\begin{equation} \label{Cfg}
\left\{C_f,\langle g[\hat{x}_i]\rangle \right\} = \frac{1}{i\hbar} 
\left\langle \left[ \hat{f} \hat{C} ,g[\hat{x}_i] \right] \right\rangle
= \frac{1}{i\hbar} \left\langle \hat{f} \left[ \hat{C}, g[\hat{x}_i] 
\right] + \left[ \hat{f}, g[\hat{x}_i] \right]\hat{C} \right\rangle
= \frac{1}{i\hbar} \left\langle \left[ \hat{f}, g[\hat{x}_i] \right] 
\hat{C} \right\rangle
\end{equation}
which vanishes on the constraint surface. This means that the
variables $F^{a_1, a_2,\ldots a_{2N}; 0, 0}$\ are both unconstrained
and unaffected by the gauge flows. They can be used to construct the
quantum variables corresponding to precisely $N$\ canonical pairs, so
that we have \emph{at least} the correct number of physical degrees of
freedom. Finally we show that the variables $F^{a_1, a_2,\ldots
a_{2N}; b, 0}$, $b\neq0$\ are \emph{pure gauge}
\begin{eqnarray}
\{C_f, F^{a_1, a_2,\ldots a_{2N}; b, 0}\} &=& \frac{1}{i\hbar} 
\left\langle \left[ \hat{f} \hat{C}, (\hat{x}_1)^{a_1} \cdots
(\hat{x}_{2N})^{a_{2N}}\hat{p}^b \right] \right\rangle \nonumber\\
&=& \frac{1}{i\hbar} \left\langle \left[ \hat{f}, (\hat{x}_1)^{a_1} 
\cdots (\hat{x}_{2N})^{a_{2N}}\hat{p}^b \right] \hat{C} + i\hbar b 
 \hat{f} (\hat{x}_1)^{a_1} \cdots (\hat{x}_{2N})^{a_{2N}}\hat{p}^{b-1} 
\right\rangle \nonumber\\
&\approx& b \left\langle \hat{f} (\hat{x}_1)^{a_1} \cdots 
(\hat{x}_{2N})^{a_{2N}}\hat{p}^{b-1} \right\rangle
\end{eqnarray}
where ``$\approx$'' denotes equality on the constraint
surface. Substituting a constraint such that $\hat{f}=g[\hat{x}_i]
\hat{C}^{b-1}$
\[
\left\{C_{gC^{b-1}}, F^{a_1, a_2,\ldots a_{2N}; b, 0} \right\} 
\approx b\left\langle\ g[\hat{x}_i] \left((\hat{x}_1)^{a_1} \cdots 
(\hat{x}_{2N})^{a_{2N}}\right) \hat{C}^{b-1}\hat{p}^{b-1} \right\rangle
\]
and commuting all the $\hat{C}$ to the right one by one, such that
$\hat{C}^{b-1}\hat{p}^{b-1} = (b-1)!(i\hbar)^{b-1}+\cdots$ up to operators 
of the form $\hat{A}\hat{C}$, we have
\begin{equation}
\left\{C_{gC^{b-1}}, F^{a_1, a_2,\ldots a_{2N}; b, 0} \right\} \approx
b!(i\hbar)^{b-1}\left\langle\ g[\hat{x}_i] \left( (\hat{x}_1)^{a_1}
\cdots (\hat{x}_{2N})^{a_{2N}} \right) \right\rangle\,.
\label{eq:F_gauge_flow}
\end{equation}
Since the right-hand side is a gauge independent function,
(\ref{eq:F_gauge_flow}) tells us that it is impossible to pick a gauge
where all of the flows on a given variable $F^{a_1, a_2,\ldots a_{2N};
b, 0}$\ vanish, in this sense we refer to all such variables as
\emph{pure gauge}.

To summarize: using an alternative set of variables $F^{a_1,
a_2,\ldots a_{2N}; b, c}$\ defined in Eq.~(\ref{eq:Fvariables}) we
find that constraints become $F^{a_1, a_2,\ldots a_{2N}; b,
c}\approx0$, $c\neq0$; the variables $F^{a_1, a_2,\ldots a_{2N}; b,
0}$, $b\neq0$ are pure gauge, which leaves the gauge invariant and
unconstrained physical variables $F^{a_1, a_2,\ldots a_{2N}; 0,
0}$. These may then be used to determine directly the physical quantum
variables $G^{a_1, \ldots, a_{2N}; 0, 0}$ defined in
Eq.~(\ref{eq:Qvariables}). Thus, for a linear constraint a correct
reduction in the degrees of freedom is achieved by applying
constraints of the form $C_f=\langle \hat{f} \hat{C} \rangle$\
(polynomial in the canonical variables), as can be directly observed
order by order in the quantum variables. Locally, our procedure of
effective constraints is complete and consistent since any irreducible
constraint can locally be chosen as a canonical coordinate.

\subsection{Generating functional}
\label{s:Generate}

More generally, one can work with a generating functional of all
constraints with polynomial-type multipliers, which can then be extended to
arbitrary constraints including non-linear ones.

To elaborate, we return to a single canonical pair and denote basic
operators as $(\hat{x}^i)_{i=1,2}=(\hat{q},\hat{p})$ such that they
satisfy the Heisenberg algebra $[\hat x^i, \hat
x^j]=i\hbar\epsilon^{ij}$, where $\epsilon^{ij}$ are the components of
the non-degenerate antisymmetric tensor with $\epsilon^{12}=1$. We
assume that there is a Weyl ordered constraint operator $C(\hat x^i)$
obtained by inserting the basic operators in the classical constraint
and then Weyl ordering. We can generate the Weyl ordered form of all
quantum constraints and their algebra through use of a generating
functional, defining $C_\alpha(\hat
x^i):=e^{\frac{i}{\hbar}\alpha_i\cdot\hat x^i} C(\hat x^i)$ for all
$\alpha_i\in\R$, which turn out to form a closed algebra. It is clear
that $\langle C_{\alpha}(\hat x^i)\rangle=0$ for physical states, and
thus we have a specific class of infinitely many quantum
constraints. This class includes polynomials as multipliers which
arise from
\[
 \langle\hat{q}^a\hat{p}^b\hat{C}\rangle\propto
\left. \left(\frac{\partial^{a+b}}{\partial\alpha_1^a\partial\alpha_2^b}
\langle C_\alpha(\hat x^i)\rangle\right)\right|_{\alpha=0}
\]
in specific orderings as Weyl ordered versions of
$\hat{q}^a\hat{p}^b\hat{C}$ such that expectation values remain zero
in physical states because $\langle C_{\alpha}(\hat{x}^i)\rangle=0$
for all $\alpha$. From Sec.~\ref{s:Closure} one may suspect that this
system is not closed, but closure does turn out to be realized.  To
establish this, we provide several auxiliary calculations. First, we
have
\begin{eqnarray}
\nonumber[\hat x^{(i_1}\cdots\hat x^{i_{n})},\hat
x^{j}]_+ &=&\frac{1}{2}\delta_{(j_1}^{i_1}\cdots\delta_{j_n)}^{i_n}
\left(\hat x^{j_1}\cdots\hat x^{j_{n}}\hat x^j+ \hat x^j\hat
x^{j_1}\cdots\hat x^{j_{n}}\right)\\\nonumber
&=&\frac{1}{2(n+1)}\delta_{(j_1}^{i_1}\cdots\delta_{j_n)}^{i_n}
\left[2\sum_{r=0}^n
\hat x^{j_1}\cdots\hat x^{j_{r}}\hat x^j\hat x^{j_{r+1}}\cdots\hat
x^{j_{n}} \right.\\\nonumber && +\sum_{r=1}^n
i\hbar(n+1-r)\epsilon^{jj_{r}}\hat x^{j_1}\cdots\hat x^{j_{r-1}}\hat
x^{j_{r+1}}\cdots\hat x^{j_{n}}\\\nonumber && +\left.\sum_{r=1}^n
i\hbar(n+1-r)\epsilon^{j_{n-r}j}\hat x^{j_1}\cdots\hat
x^{j_{n-r-1}}\hat x^{j_{n-r+1}}\cdots\hat x^{j_{n}}\right]\\ &=&\hat
x^{(i_1}\cdots\hat x^{i_{n}}\hat x^{j)}\,. \label{Weyl}
\end{eqnarray}
Thus, the anticommutator of a Weyl ordered operator with a basic
operator is also Weyl ordered.

From Baker--Campbell--Hausdorff identities it follows that
$e^{\frac{i}{\hbar}\alpha_i\cdot\hat x^i}$ acts as a displacement
operator
\begin{equation}
 e^{\frac{i}{\hbar}\alpha_i\cdot\hat x^i}\hat
 x^je^{-\frac{i}{\hbar}\alpha_i\cdot\hat x^i}=\hat
x^j+\epsilon^{ji}\alpha_i\,.
\end{equation}
This also shows the algebra
of these operators:
\begin{equation}
 e^{\frac{i}{\hbar}\alpha_i\cdot\hat
 x^i}e^{\frac{i}{\hbar}\beta_i\cdot\hat x^i}=
 e^{\frac{i}{\hbar}\alpha_i\cdot\hat
x^i+\frac{i}{\hbar}\beta_i\cdot\hat
x^i-\frac{1}{2\hbar^2}[\alpha\cdot\hat x,\beta\cdot\hat
x]}=e^{\frac{i}{\hbar}(\alpha_i+\beta_i)\cdot\hat
x^i}e^{-\frac{i}{2\hbar}\alpha_i\epsilon^{ij}\beta_j}\,.
\end{equation}

With this, one can realize the operator $C_\alpha(\hat
x^i)$ as
\begin{eqnarray}
C_\alpha(\hat x^i)&:=&e^{\frac{i}{\hbar}\alpha_i\hat x^i} C(\hat x^i)=
e^{\frac{i}{2\hbar}\alpha_i\hat x^i} C(\hat
x^i+{\textstyle\frac{1}{2}}
\epsilon^{ij}\alpha_j)e^{\frac{i}{2\hbar}\alpha_i\hat
x^i}\nonumber\\&=&\sum_{n=0}^{\infty}\sum_{m=0}^n\frac{1}{2^n\hbar^nn!}
{n\choose m} (i\alpha\cdot\hat
x)^m C(\hat x^i+{\textstyle\frac{1}{2}}\epsilon^{ij}\alpha_j)(i\alpha\cdot\hat
x)^{n-m}\nonumber\\&=&
\sum_{n=0}^{\infty}\frac{1}{\hbar^nn!} [i\alpha\cdot\hat x,C(\hat
x^i+{\textstyle\frac{1}{2}}\epsilon^{ij}\alpha_j)]_{+n}\,,
\end{eqnarray}
which is manifestly Weyl ordered due to (\ref{Weyl}). Here, we use the
iterative definition $[\hat{A},\hat{C}]_{+0}:=\hat{C}$ and
$[\hat{A},\hat{C}]_{+n}:=[\hat{A},[\hat{A},\hat{C}]]_{+(n-1)}$.

Finally, the algebra of constraints is
\begin{eqnarray}
\left[C_\alpha(\hat x^i),C_\beta(\hat
x^i)\right]&=&\left(e^{\frac{i}{\hbar}\alpha_i\hat x^i} C(\hat
x^i)e^{\frac{i}{\hbar}\beta_i\hat x^i}-e^{\frac{i}{\hbar}\beta_i\hat
x^i} C(\hat x^i)e^{\frac{i}{\hbar}\alpha_i\hat x^i}\right) C(\hat x^i)
\\\nonumber&=&\left(e^{\frac{i}{2\hbar}\alpha_i\hat x^i} C(\hat
x^i+{\textstyle\frac{1}{2}}\epsilon^{ij}\alpha_j)e^{\frac{i}{\hbar}\beta_i(\hat
x^i+\frac{1}{2}\epsilon^{ij}\alpha_j)}e^{\frac{i}{2\hbar}\alpha_i\hat
x^i}\right.\\\nonumber&&\left.-e^{\frac{i}{2\hbar}\alpha_i\hat
x^i}e^{\frac{i}{\hbar}\beta_i(\hat
x^i-\frac{1}{2}\epsilon^{ij}\alpha_j)} C(\hat
x^i-{\textstyle\frac{1}{2}}\epsilon^{ij}\alpha_j)e^{\frac{i}{2\hbar}\alpha_i\hat
x^i}\right)C(\hat x^i)
\\\nonumber&=&\left(e^{\frac{i}{\hbar}\beta_i\frac{1}{2}
\epsilon^{ij}\alpha_j}e^{\frac{i}{2\hbar}(\alpha_i+\beta_i)\hat x^i}
C(\hat x^i+{\textstyle\frac{1}{2}}\epsilon^{ij}(\alpha_j-\beta_j))
e^{\frac{i}{2\hbar}(\alpha_i+\beta_i)\hat x^i}\right.\\\nonumber&&
\left.-e^{-\frac{i}{\hbar}\beta_i\frac{1}{2}\epsilon^{ij}\alpha_j}
e^{\frac{i}{2\hbar}(\alpha_i+\beta_i)\hat x^i} C(\hat
x^i-{\textstyle\frac{1}{2}}\epsilon^{ij}(\alpha_j-\beta_j))
e^{\frac{i}{2\hbar}(\alpha_i+\beta_i)\hat x^i}\right)C(\hat x^i)
\\
\end{eqnarray}
and thus
\begin{eqnarray}
\left[C_\alpha(\hat x^i),C_\beta(\hat
x^i)\right]&=&\left[e^{\frac{i}{2\hbar}\beta_i\epsilon^{ij}\alpha_j}C
(\hat{x}^i
+{\textstyle
\frac{1}{2}}\epsilon^{ij}(\alpha_j-\beta_j))\right.\\\nonumber&&
\left.-e^{-\frac{i}{2\hbar}\beta_i\epsilon^{ij}\alpha_j}
C(\hat{x}^i-
{\textstyle\frac{1}{2}}\epsilon^{ij}(\alpha_j-\beta_j))\right]
C_{\alpha+\beta}(\hat x^i)\,.
\end{eqnarray}
This produces a closed set of Weyl ordered and thus real effective
constraints, which is uncountable. There are closed subsets obtained
by allowing $\alpha_i$ to take values only in a lattice in phase
space, but in this case the completeness issue becomes more difficult
to address. Moreover, the $C_{\alpha}$ may be difficult to compute in
specific examples.  At this stage, we turn to a discussion of specific
examples based on polynomial multipliers in quantum constraints,
rather than providing further general properties of Weyl ordered
effective constraints.

\section{Linear examples}
\lb{Examples} 

Given that the precise implementation of a set of
quantum constraints depends on the form of the constrained system, we
illustrate typical properties by examples, starting with linear
ones.

\subsection{A canonical variable as constraint: $\hat{C}=\hat{q}$}
\label{S:Cq}

From $C^{(n)}=0$ we obtain that all quantum variables $G^{q^n}$ are
constrained to vanish, in addition to $C_Q=q$ itself. Moreover,
$C_q^{(n)}=C^{(n+1)}$ is already included, and $C_p^{(n)}=
\langle\hat{p}\hat{q}^n\rangle$ provides a closed set of
constraints. In fact, computing commutators does not add any new
constraints and we already have a closed, first class system which
suffices to discuss moments up to second order. At higher orders, also
$C_{p^m}^{(n)}= \langle\hat{p}^m\hat{q}^n\rangle$ must be included.

In this example, it is feasible to work with the symmetrically ordered
quantum variables since there is an obvious quantum observable
$\hat{q}$ commuting with the constraint. For instance, quantum
variables $G^{C^nq}$ and $G^{C^np}$ form a closed set of constraints
as shown by the previous calculations.  The first class nature of this
system can directly be verified from the Poisson relations
(\ref{Gbrackets}).  For $b=d=0$ we obviously have
$\{G^{a,0},G^{c,0}\}=0$, for $b=0$ and $d=1$ we have
$\{G^{a,0},G^{c,1}\}= a(G^{a+c-1,0}- G^{a-1,0}G^{c,0})\approx 0$ and
for $b=d=1$, $\{G^{a,1},G^{c,1}\}= (a-c) G^{a+c-1,1}-
aG^{a-1,1}G^{c,0}+c G^{a,0}G^{c-1,1}\approx 0$.

To discuss moments up to second order, constraints with at most a
single power of $p$ are needed. These constraints are in fact
equivalent to constraints given by quantum variables due to
\begin{eqnarray}
 G^{q^n} &=& \langle(\hat{q}-q)^n\rangle = \sum_{j=0}^n
{n\choose j}
(-1)^jq^j\langle\hat{q}^{n-j}\rangle
 = \sum_{j=0}^{n-1}{n\choose j}
(-1)^j q^jC^{(n-j)} +(-1)^nq^n\\
 G^{q^np} &=& \frac{1}{n+1}\langle (\hat{q}-q)^n(\hat{p}-p)+
(\hat{q}-q)^{n-1}(\hat{p}-p)(\hat{q}-q)+ \cdots+
(\hat{p}-p)(\hat{q}-q)^n\rangle\nonumber\\
&=& \frac{1}{n+1}\langle (n+1)(\hat{p}-p)(\hat{q}-q)^n+
{\textstyle\frac{1}{2}}in(n+1)\hbar(\hat{q}-q)^{n-1}\rangle\nonumber\\
&=& \langle\hat{p}\hat{q}^n\rangle -p\langle(\hat{q}-q)^n\rangle+
\sum_{j=1}^n{n\choose j}
(-1)^j q^j\langle\hat{p}\hat{q}^{n-j}\rangle+
\frac{1}{2}in\hbar \langle(\hat{q}-q)^{n-1}\rangle\nonumber\\
&=& C_p^{(n)}-pG^{q^n}+ \sum_{j=1}^{n-1}
{n\choose j}
(-1)^jq^iC_p^{(n-j)} +(-1)^nq^np+ \frac{1}{2}i\hbar nG^{q^{n-1}}\,.
\end{eqnarray}
This describes a one-to-one mapping from
$\{C^{(n)},C_p^{(m-1)}\}_{n,m\in{\mathbb N}}$ to
$\{G^{q^n},G^{q^mp}\}_{n,m\in{\mathbb N}}$ which provides specific
examples of the relation between (\ref{eq:Qvariables}) and
(\ref{eq:Fvariables}) in Sec.~\ref{S:Number}.  Thus, the constraint
surface as well as the gauge flow can be analyzed using quantum
variables. For this type of classical constraint, reordering will only
lead to either a constant or to terms depending on quantum variables
defined without reference to $\hat{p}$.  Since these are already
included in the set of constraints and a constant does not matter for
generating canonical transformations, they can be eliminated when
computing the gauge flow. Note, however, that there is a constant term
$\frac{1}{2}i\hbar$ in $G^{q^np}$ for $n=1$ which will play an
important role in determining the constraint surface. The fact that
constraints are complex valued does not pose a problem for the gauge
flow since imaginary contributions come only with coefficients which
are (real) constraints themselves and thus vanish weakly, or are
constant and thus irrelevant for the flow.

Also the gauge flow up to second order generated by the quantum
constraints can be computed using quantum variables such as $G^{q^n}$
and $G^{q^np}$ rather then the non-symmetric version.  For the moments
of different orders, we then have the following constraints and gauge
transformations. (i) Expectation values: one constraint $q\approx0$
generating one gauge transformation $p\mapsto p+\lambda_1$. (ii)
Fluctuations: two constraints $G^{qq} \approx 0$ and $G^{qp}\approx
{\rm const}$, generating gauge transformations $G^{pp}\mapsto
G^{pp}+4\lambda_2 G^{qp}$ and $G^{pp}\mapsto G^{pp}(1+2\lambda_2)$,
respectively. As we will see in Eq.~(\ref{Gqp}) below, $G^{qp}$ is
non-zero on the constraint surface, such that $G^{pp}$ is completely
gauge. (iii) Higher moments: at each order, we have constraints
$C_{p^m}^{(n-m)}$ with $m<n$ and only $G^{p^n}$ is left to be removed
by gauge generated e.g.\ by $G^{q^n}$. This confirms the counting of
Sec.~\ref{S:Number}. Moreover, higher order constraints generate a
gauge flow which also affects fluctuations, in particular
$G^{pp}$. Thus, to second order we see that two moments are eliminated
by quantum constraints while the remaining one is gauge.  In this way,
the quantum variables are eliminated completely either by constraints
or by being pure gauge. (Moments such as $G^{qp}$ were not included in
the counting argument of Sec.~\ref{S:Number} in the context of the
dimension of the gauge flow to be factored out. Here, in fact, we
verify that the flow generated by $G^{qq}$ suffices to factor out all
remaining quantum variables to second order.)

This example also illustrates nicely the role of imaginary
contributions to the constraints from the perspective of the
kinematical inner product. The constraint
$C_p^{(1)}=\langle\hat{p}\hat{q}\rangle=0$  implies that
\begin{equation} \label{Gqp}
 G^{qp}=\frac{1}{2}\langle\hat{q}\hat{p}+\hat{p}\hat{q}\rangle-qp
 =\langle\hat{p}\hat{q}\rangle-qp+\frac{1}{2}i\hbar\approx
 \frac{1}{2}i\hbar
\end{equation}
must be imaginary. From the point of view of the kinematical inner
product this seems problematic since we are taking the expectation
value of a symmetrically ordered product of self-adjoint
operators. However, the inner product of the kinematical Hilbert space
is only auxiliary, and from our perspective not even necessary to
specify. Then, an imaginary value (\ref{Gqp}) of some kinematical
quantum variables has a big advantage: it allows us to formulate the
quantum constrained system without violating uncertainty
relations. For an unconstrained system, we have the generalized
uncertainty relation
\begin{equation}
 G^{qq}G^{pp}-(G^{qp})^2 \geq \frac{1}{4}\hbar^2\,.
\end{equation}
This relation, which is important for an analysis of coherent states,
would be violated had we worked with real quantum constraints
$G^{qq}\approx 0\approx G^{qp}$ instead of
$(C^{(2)},C_p^{(1)})$. Again, this is not problematic because the
uncertainty relation is formulated with respect to the kinematical
inner product, which may change. Still, the uncertainty relations are
useful to construct coherent states and it is often helpful to have
them at ones disposal. They can be formulated without using
self-adjointness, but this would require one to treat $\hat{q}$,
$\hat{p}$ as well as $\hat{q}^{\dagger}$ and $\hat{p}^{\dagger}$ as
independent such that their commutators (needed on the right hand side
of an uncertainty relation) are unknown.  The imaginary value of
$G^{qp}$ obtained with our definition of the quantum constraints, on
the other hand, allows us to implement the constraints in a way
respecting the standard uncertainty relation:
$-(G^{qp})^2=\frac{1}{4}\hbar^2$ from (\ref{Gqp}) saturates the
relation.

\subsection{Discrete momentum as constraint: $\hat{C}=\hat{p}$ on a
circle}
\label{s:circle}

We now assume classical phase space variables $\phi\in S^1$ with
momentum $p$. This requires a non-canonical basic algebra generated by
the operators $\hat{p}$, $\widehat{\sin\phi}$ and $\widehat{\cos\phi}$
with
\begin{equation} \label{circlealg}
 [\widehat{\sin\phi},\hat{p}] = \widehat{\cos\phi} \quad,\quad
 [\widehat{\cos\phi},\hat{p}] = -\widehat{\sin\phi}\,.
\end{equation}
This example can also be seen as a model for isotropic loop quantum
cosmology and gravity \cite{IsoCosmo,Bohr,LivRev}.

The constraint operator $\hat{C}=\hat{p}$ implies the presence of
quantum constraints $C_Q=p$ as well as $C_p^{(n-1)}\approx
G^{p^n}$. This is not sufficient to remove all quantum variables by
constraints or gauge, and we need to include quantum constraints
referring to $\phi$. Unlike in Sec.~\ref{S:Cq}, we cannot take
$f=\phi$ because there is no operator for $\phi$. If we choose
$C_{\sin\phi}^{(n)}$ as starting point, the requirement of a closed
set of constraints generates $C_{1\cdot[p,\sin\phi]}^{(n)} =
-C_{\cos\phi}^{(n)}$. Taken together, those constraints generate
$C_{\sin\phi[p,\cos\phi]}^{(n)}=C_{\sin^2\phi}^{(n)}$,
$C_{\sin\phi[p,\sin\phi]}^{(n)}=-C_{\sin\phi\cos\phi}^{(n)}$ and
$C_{\cos\phi[p,\sin\phi]}^{(n)}=-C_{\cos^2\phi}^{(n)}$, i.e.\ all
quantum constraints $C_{f(\phi)}^{(n)}$ with a function $f$ depending
on $\phi$ polynomially of second degree through $\sin\phi$ and
$\cos\phi$. Iterating the procedure results in a closed set of
constraints $p$, $G^{p^n}$ and $C_{P(\sin\phi,\cos\phi)}^{(n)}$ with
arbitrary polynomials $P(x,y)$.

In this case, we have independent uncertainty relations for each pair
of self-adjoint operators. Relevant for consistency with the
constraints is the relation
\[
 G^{pp}G^{\cos\phi\cos\phi}-(G^{p\cos\phi})^2\geq
\frac{1}{4}\hbar^2\langle\widehat{\sin\phi}\rangle^2
\]
and its obvious analog exchanging $\cos\phi$ and $\sin\phi$. Also
here, one can see as before that the imaginary part of
$G^{p\cos\phi}=C_{\cos\phi}^{(1)}-p\cos\phi
+\frac{1}{2}i\hbar\sin\phi\approx \frac{1}{2}i\hbar\sin\phi$ allows
one to respect the uncertainty relation even though
$G^{pp}\approx0$. 

Note that this is similar to the previous example,
although now zero being in the discrete spectrum of $\hat{p}$ would
allow one to use a physical Hilbert space as a subspace of the
kinematical one whose reality conditions could thus be preserved. If
this is done, $G^{p\cos\phi}$ must be real even kinematically
because the kinematical inner product determines the physical one
just by restriction. The uncertainty relation in this example turns
out to be respected automatically, even for real kinematical quantum
variables, because the algebra (\ref{circlealg}) of operators
implies that $\langle\widehat{\sin\phi}\rangle=
\langle[\widehat{\cos\phi},\hat{p}]\rangle=0$ in physical states.
Moreover, $G^{p\cos\phi}\approx \frac{1}{2}i\hbar\sin\phi\approx 0$
turns out to be real on the constraint surface, after all.

Alternatively, if one knows that the constraint is represented as a
self-adjoint operator with zero in the discrete part of its spectrum,
the same relations can be recovered by appealing directly to the
existence of creation and annihilation operators which map zero
eigenstates of the constraint to other states in the kinematical
Hilbert space. For these operators to exist, the physical Hilbert
space must indeed be a subspace of the kinematical Hilbert space
(given by zero eigenstates of the constraint operator and the inner
product on those states) such that this argument explicitly refers to
the discrete spectrum case only. Using this information about the
quantum representation makes it possible to do the reduction of
effective constraints without introducing complex-valued kinematical
quantum variables. Indeed, in our case $\hat{a}^{\dagger} =
\widehat{\cos\phi} + i\widehat{\sin\phi}$ and $\hat{a} =
\widehat{\cos\phi} - i\widehat{\sin\phi}$, respectively, raise and
lower the discrete eigenvalues of $\hat{p}$ represented on the Hilbert
space $L^2(S^1,\md\phi)$. For any eigenstate of $\hat{p}$, then,
$\langle \hat{a}^{\dagger} \rangle= \langle\widehat{\cos\phi}\rangle+
i\langle\widehat{\sin\phi}\rangle = 0$ and $\langle \hat{a} \rangle =
\langle\widehat{\cos\phi}\rangle-
i\langle\widehat{\sin\phi}\rangle=0$. Thus, we again derive that the
right hand side of uncertainty relations vanishes in physical states,
making real-valued kinematical quantum variables consistent. Moreover,
this example shows that for a constraint with zero in the discrete
part of its spectrum, additional constraints follow which can be used
to eliminate variables which in the general effective treatment appear
as gauge. In fact, all moments involving $\sin\phi$ or $\cos\phi$ are
constrained to vanish if
$\langle\hat{a}^n\rangle=0=\langle(\hat{a}^{\dagger})^m\rangle$
is used for physical states. In this case, no gauge flow is necessary
to factor out these moments, but in contrast to the gauge flow itself
this can only be seen based on representation properties.

Using complex valued kinematical quantum variables turns out to be
more general and applicable to constraints with zero in the discrete
or continuous spectrum. For systems with zero in the discrete
spectrum, this can be avoided but requires one to refer explicitly to
properties of the quantum representation or the operator algebra.

\subsection{Two component system with constraint:
  $\hat{C}=\hat{p}_1-\hat{p}$}

As an example which can be interpreted as a parameterized version of
an unconstrained system, we consider a system with a $4$-dimensional
phase space and phase space coordinates $(q,p;q_1,p_1)$. The system is
governed by a linear constraint \be C_Q=p_1- p\ .  \ee The classical
constraint can, of course, be transformed canonically to a constraint
which is identical to one of the phase space coordinates since
$(\frac{1}{2}(q_1-q),C;\frac{1}{2}(q_1+q),p_1+p)$ forms a system of
canonical coordinates and momenta containing $C=p_1-p$. Moreover, the
transformation is linear and can easily be taken over to the quantum
level as a unitary transformation. The orders of moments do not mix
under such a linear transformation, and thus the arguments put forward
in Sec.~\ref{S:Cq} can directly be used to conclude that the system
discussed here is consistent and complete. Nevertheless, it is
instructive to look at details of the procedure without doing such a
transformation, which will serve as a guide for more complicated
cases.

Expectation values satisfy the classical gauge transformations
\begin{equation}\label{ppgauge}
 -\dot{q}=1=\dot{q}_1\quad,\quad \dot{p}=0=\dot{p}_1\,.
\end{equation}
At this point, we recall that there are no reality or positivity
conditions for the kinematical quantum variables (\ref{QuantVars2}) as
they appear before solving any constraints.  Their gauge
transformations are \be\label{ppgaugeG} \dot{G}^{a,b}_{c,d}=0\ , \ee
where \be G^{a,b}_{c,d}=\langle(\hat p-p)^a(\hat q-q)^b(\hat
p_1-p_1)^c(\hat q_1-q_1)^d\rangle_{\rm Weyl}\, .\ee Even though these
variables remain constant, as do those of the deparameterized system,
here we have additional moments compared to an unconstrained canonical
pair: solving the constraints has to eliminate all quantum variables
with respect to one canonical pair, but also cross-correlations to the
unconstrained pair. These cannot all be set to zero simultaneously due
to the uncertainty relations --- but they may be chosen to satisfy
minimal uncertainty.

\subsubsection{Constraints}

In addition to gauge transformations (\ref{ppgauge}) and
(\ref{ppgaugeG}) generated by the principal quantum constraint
$C_Q=C^{(1)}$, the system is subject to further constraints and their
gauge transformations. As explained above, the quantum constraints
have to form a complete, first class set. Such a set is given by
\ben
C^{(n)}&=&\Mmkl p_1^kp^{\ell}G^{n-m-\ell,0}_{m-k,0}
\\
C^{(n)}_q&=&\Mmkl
p_1^kp^{\ell}\left(G^{n-m-\ell,1}_{m-k,0}
-\frac{i\hbar}{2}(n-m-\ell) G^{n-m-\ell-1,0}_{m-k,0}
\right)
\\ C^{(n)}_p&=&\Mmkl p_1^kp^{\ell}\left(pG^{n-m-\ell,0}_{m-k,0}+
G^{n-m-\ell+1,0}_{m-k,0}\right)
\\
C^{(n)}_{p_1}&=&\Mmkl p_1^kp^{\ell}\left(p_tG^{n-m-\ell,0}_{m-k,0}+
G^{n-m-\ell,0}_{m-k+1,0}\right)
\\
C^{(n)}_{q_1}&=&\Mmkl
p_1^kp^{\ell}\left(G^{n-m-\ell,0}_{m-k,1}
+\frac{i\hbar}{2}(m-k)G^{n-m-\ell,0}_{m-k-1,0}
\right)\ .  
\een
These constraints are accompanied by analogous
expressions involving polynomial factors of the basic operators, which
we will not be using to the orders considered here. We thus solve our
constraints as given to the required orders and determine the gauge
orbits they generate.

At this point, a further choice arises: we need to determine which
variables we want to solve in terms of others which are to be kept
free. This is related to the choice of time in a deparametrization
procedure. Here, we view $q_1$ as the time variable which is demoted
from a physical variable to the status of an evolution parameter, and
thus $H=p$ will be the Hamiltonian generating evolution in this
time. Notice that time is chosen after quantization when dealing with
effective constraints. (For our linear constraint, of course, the
roles of the two canonical pairs can be exchanged, with $q$ playing
the role of time.)

Classically, it is then straightforward to solve the constraint and
discuss gauge, and the same applies to expectation values in the
quantum theory. The discussion of quantum variables is, however,
non-trivial and is therefore presented here in some detail for second
order moments. Having made a choice of time, a complete
deparametrization requires that all quantum variables of the form
$G^{a_1,b_1}_{a_2,b_2}$ with $a_2\not=0$ or $b_2\not=0$ be completely
constrained or removed by gauge. Only quantum variables
$G^{a,b}_{0,0}$ are allowed to remain free, and must do so without any
further restrictions.  To second order, the deparametrized system has
$2+3=5$ variables; the parametrized theory has $4+10=14$. We begin by
eliminating quantum variables in favor of the variables associated
with the canonical pair $(q_1,p_1)$ only.  From the fact that, on the
one hand, $G_{p_1p_1}$, $G_{p_1q_1}$ and $G_{q_1q_1}$ should satisfy
the uncertainty relations and thus cannot all vanish but, on the
other hand, are not present in the unconstrained system, we expect at
least one of them to be removed by gauge.

At second order,\footnote{The moment expansion is formalized in
Sec.~\ref{s:ConsCons}.} i.e.\ keeping only second order moments as
well as terms linear in $\hbar$, the constraints form a closed and
complete system given by
\ben
C^{(n)}\vert_{N=2}&=&c_n+d_n\left(G^{pp}+G_{p_1p_1}-2G^p_{p_1}\right)\\
C^{(n)}_p\vert_{N=2}&=&a_n\left(G^{pq}-G^p_{p_1}-
\frac{\mathrm{i}\hbar}{2}\right)+b_n\left(G^{pp}-2G^p_{p_1}+G_{p_1p_1}\right)\\
C^{(n)}_q\vert_{N=2}&=&a_n\left(G^{pp}-G^p_{p_1}\right)\\
C^{(n)}_{p_1}\vert_{N=2}&=&a_n\left(G^p_{p_1}-G_{p_1p_1}\right)\\
C^{(n)}_{q_1}\vert_{N=2}&=&a_n\left(G^p_{q_1}-G_{p_1q_1}-
\frac{\mathrm{i}\hbar}{2}\right)+
c_n\left(G_{p_1p_1}-2G^p_{p_1}+2c_nG^{pp}\right)\ ,
\een
where
\ben
a_n&=&a_n(p_1,q)\equiv -n(C^{(1)})^{n-1}\quad,\quad
b_n=b_n(p_1,q)\equiv\frac{\mathrm{i}\hbar}{2}\frac{n}{2}(n-1)(n-2)
(C^{(1)})^{n-3}\\
c_n&=&c_n(p_1,q)\equiv (C^{(1)})^{n}\quad,\quad
d_n=d_n(p_1,q)\equiv\frac{n}{2}(n-1)(C^{(1)})^{n-2}\ ,
\een
and $C^{(1)}=p_1-p$ is the linear constraint which in this case is
identical with the classical constraint.

Due to the fact that the prefactors in the constraint equations
contain $C^{(1)}$, we find non-trivial constraints only when the
exponent of $C^{(1)}$ vanishes. This happens for $a_1$, $b_3$ and
$d_2$, while $c_n$ vanishes for all $n$. For higher $n$ no additional
constraints arise. Constraints arising for $n=2,3$ turn out to be
linear combinations of the constraints arising for $n=1$.  Therefore
we find for the second order system only five independent constraints:
$C^{(1)}\vert_{N=2}=p_1-p$ and
\ben
C^{(1)}_q\vert_{N=2}&=&-\frac{\mathrm{i}\hbar}{2}-G^{qp}+G^q_{p_1}\quad,\quad
C^{(1)}_p\vert_{N=2}=G^p_{p_1}-G^{pp}\\
C^{(1)}_{p_1}\vert_{N=2}&=&G_{p_1p_1}-G^p_{p_1}\quad,\quad
C^{(1)}_{q_1}\vert_{N=2}=\frac{\mathrm{i}\hbar}{2}-G^p_{q_1}+G_{q_1p_1}\ .
\een
From these equations it is already obvious that four second order
moments referring to $q_1$ or $p_1$ can be eliminated through the use
of constraints. In addition to $p_1=p$ for expectation values, these
are
\begin{equation}
 G^q_{p_1}\approx \frac{1}{2}i\hbar+G^{qp} \quad,\quad
 G^p_{p_1}\approx G^{pp}\quad,\quad G_{p_1p_1} \approx
 G^p_{p_1}\approx G^{pp}
\end{equation}
as well as
\begin{equation}
 G^p_{q_1}\approx \frac{1}{2}i\hbar+ G_{q_1p_1}
\end{equation}
which is not yet completely expressed in terms of moments only of
$(q,p)$. The remaining moments of $(q_1,p_1)$ are not constrained at
all, and thus must be eliminated by gauge transformations.  To
summarize, three expectation values are left unconstrained, one of
which should be unphysical; six second-order variables are
unconstrained, three of which should be unphysical. Notice that there
is no contradiction to the fact that we have four weakly commuting
(and independent) constraints but expect only three variables to be
removed by gauge. These are constraints on the space of second order
moments, which, in this truncation, as noted before do not have a
non-degenerate Poisson bracket (although the space of all moments has
a non-degenerate symplectic structure). Weak commutation then does not
imply first class nature in the traditional sense (see e.g.\
\cite{brackets}), and four weakly commuting constraints may declare
less than four variables as gauge. While the constraints as
functionals are independent, their gauge flows may be linearly
dependent.

\subsubsection{Observables}

To explicitly account for the unphysical degrees of freedom, we
consider the gauge transformations generated by the constraints.  The
quantum constraint $p_1-p \approx 0$ produces a flow on the
expectation values only, which agrees with the classical flow
(\ref{ppgauge}).  The second-order constraints, produce no
(independent)\footnote{The parts of the second-order constraints
proportional to $C^{(1)}$ that have been discarded can also be ignored
when computing the flows generated on the constraint surface, as the
missing contributions are proportional to the gauge flow associated
with $C^{(1)}$. This is true in general, and extends to higher
orders.} flow on the expectation values.

Also $G^{pp}$ is gauge invariant.  For the five remaining free
second-order variables, $G^p_{p_1} - G^{pp} \approx 0$ generates a
flow (on the constraint surface):
\begin{eqnarray}
\delta G^{qp} &=& G^p_{p_1}-2G^{pp} \approx - G^{pp} \nonumber\\
\delta G^{qq} &=& 2G^q_{p_1}-4G^{qp} \approx i\hbar -
2G^{qp}\nonumber\\
\delta G_{q_1p_1} &=& G^p_{p_1}\approx G^{pp}\nonumber\\
\delta G^q_{q_1} &=& G_{q_1p_1}+G^{qp}-2G^p_{q_1} \approx
G^{qp} - G_{q_1p_1} - i\hbar \nonumber\\
\delta G_{q_1q_1}&=& G^p_{q_1} \approx i\hbar + 2G_{q_1p_1}
\label{eq:p1Cflow}
\end{eqnarray}
$G^p_{p_1} - G_{p_1 p_1} \approx 0$\ gives:
\begin{eqnarray}
&&
\delta G^{q p} \approx  G^{p p} \quad,\quad
\delta G^{q q} \approx i\hbar + 2G^{q p} \quad,\quad
\delta G_{q_1 p_1} \approx
- G^{p p}\nonumber\\
&&
\delta G^q_{q_1} \approx G_{q_1 p_1} - G^{q p} -
i\hbar \quad,\quad
\delta G_{q_1 q_1} \approx i\hbar -
2G_{q_1 p_1}\label{eq:p2Cflow}
\end{eqnarray}
$\frac{1}{2} i \hbar + G^{q p} - G^q_{p_1} \approx 0$\ gives:
\begin{eqnarray}
&&
\delta G^{q p} \approx \frac{1}{2}i\hbar + G^{q p}
\quad,\quad
\delta G^{q q} \approx 2G^{q q}
\quad,\quad
\delta G_{q_1 p_1} \approx -
\frac{1}{2}i\hbar - G^{q p}\nonumber\\
&&
\delta G^q_{q_1} \approx G^q_{q_1} - G^{q q}
\quad,\quad
\delta G_{q_1 q_1} \approx - 2G^q_{q_1}
\label{eq:q1Cflow}
\end{eqnarray}
$\frac{1}{2} i \hbar - G^p_{q_1} + G_{q_1 p_1} \approx 0$\ gives:
\begin{eqnarray}
&&
\delta G^{q p} \approx - \frac{1}{2}i\hbar -
G_{q_1 p_1} \quad,\quad
\delta G^{q q} \approx -
2G^q_{q_1} \quad,\quad
\delta G_{q_1 p_1} \approx
\frac{1}{2}i\hbar + G_{q_1 p_1}\nonumber\\
&&
\delta G^q_{q_1} \approx G^q_{q_1} - G_{q_1 q_1}
\quad,\quad
\delta G_{q_1 q_1} \approx 2G_{q_1 q_1}
\label{eq:q2Cflow}
\end{eqnarray}
All of the gauge flows obey
\begin{equation}
\delta G^{q p} = -
\delta G_{q_1 p_1} \quad,\quad
\delta G^q_{q_1} = - \frac{1}{2} \left(
\delta G^{q q} +
\delta G_{q_1 q_1} \right)\,.
\label{eq:constr_form}
\end{equation}
Thus, in addition to $A_1:=G^{pp}$ we can identify the observables
$A_2:= G^{qq}+2G^q_{q_1}+G_{q_1q_1}$ and $A_3:=
G^{qp}+G_{q_1p_1}$. They satisfy the algebra $\{A_1,A_3\}= -2A_1$,
$\{A_1,A_2\}\approx -4(A_3+\frac{1}{2}i\hbar)$, $\{A_2,A_3\} = 2A_2$
on the constraint surface which, except for the imaginary term, agrees
with the Poisson algebra expected for unconstrained quantum variables
of second order. The imaginary term can easily be absorbed into the
definition of $A_3$, which leads us to the physical quantum variables
\begin{equation} \label{physqv}
 {\cal G}^{qq}:= G^{qq}+2G^q_{q_1}+G_{q_1q_1} \quad,\quad {\cal
 G}^{pp}:= G^{pp}\quad,\quad {\cal G}^{qp} :=
 G^{qp}+G_{q_1p_1}+\frac{1}{2}i\hbar\,.
\end{equation}
They commute with all the constraints and satisfy the standard algebra
for second order moments, thus providing the correct
representation. To implement the physical inner product, we simply
demand that all the physical quantum variables be real. This means
that $G^{qp}+G_{q_1p_1}$ must have the imaginary part
$-\frac{1}{2}i\hbar$ which is possible for kinematical quantum
variables.

\subsubsection{Gauge fixing}

In fact, one can choose a gauge where all physical quantum variables
agree with the kinematical quantum variables of the pair $(q,p)$, and
kinematical quantum variables of the pair $(q_1,p_1)$ satisfy
$G_{p_1p_1}=0$ and $G_{q_1p_1}=-\frac{1}{2}i\hbar$. This choice
violates kinematical reality conditions, but it ensures physical
reality and preserves the kinematical uncertainty relation even though
one fluctuation vanishes.

Other gauge choices are possible since only $G^{qp}+G_{q_1p_1}$ is
required to have imaginary part $-\frac{1}{2}\hbar$ for real ${\cal
G}^{qp}$, which can be distributed in different ways between the two
moments. Thus, there are different choices of the kinematical reality
conditions. Such gauge choices may be related to some of the freedom
contained in choosing the kinematical Hilbert space which would
similarly affect the reality of kinematical quantum variables.

The algebra of the physical variables can be recovered without the
knowledge of their explicit form as observables, by completely fixing
the gauge degrees of freedom and using the Dirac bracket to find the
Poisson structure on the remaining free parameters. We introduce gauge
conditions $\phi_i=0$ which together with the second order constraints
define a symplectic subspace $\Sigma_{\phi}$ of the space of second
order quantum variables. Our conditions should fix the gauge freedom
entirely --- which means that the flow due to any remaining first
class constraints should vanish on $\Sigma_{\phi}$. (We recall that
the space of second order moments does not form a symplectic subspace
of the space of all moments, but it does define a Poisson manifold. In
such a situation, not all first class constraints need to be
gauge-fixed to obtain a symplectic gauge-fixing surface.)  In order to
ensure that the conditions put no restrictions on the physical degrees
of freedom, we demand that no non-trivial function of the gauge
conditions be itself gauge invariant.

The simple gauge discussed above corresponds to
$\phi_4=G_{p_1q_1}+\frac{1}{2}i\hbar=0$, $\phi_5=G_{q_1q_1}=0$ and
$\phi_6=G^q_{q_1}=0$.  Under these conditions $C^{(1)}_{q_1}$ remains
first class but has a vanishing flow (\ref{eq:q2Cflow}) on the surface
$\Sigma_{\phi}$. The other second order constraints now form a second
class system when combined with the gauge conditions. The combination
of constraints and gauge fixing conditions eliminates all second order
variables except for $G^{pp}$, $G^{qq}$ and $G^{qp}$, which therefore
parameterize $\Sigma_{\phi}$. Labeling $\phi_1=C^{(1)}_p$,
$\phi_2=C^{(1)}_q$, $\phi_3=C^{(1)}_{p_1}$, the commutator matrix
$\Delta_{ij}:=\{\phi_i, \phi_j\}$ on $\Sigma_{\phi}$,
\[
\mathbf{\Delta}|_{\Sigma_{\phi}} = \left( \begin{array}{cccccc}
0 & 0 & 0 & -G^{pp} & 0 & \frac{1}{2}i\hbar - G^{qp} \\
0 & 0 & 0 & \frac{1}{2}i\hbar + G^{qp} & 0 & G^{qq} \\
0 & 0 & 0 & G^{qq} & -2i\hbar & \frac{1}{2}i\hbar + G^{qp} \\
G^{pp} & -\frac{1}{2}i\hbar - G^{qp} & -G^{qq} & 0 & 0 & 0 \\
0 & 0 & 2i\hbar & 0 & 0 & 0 \\
G^{qp} - \frac{1}{2}i\hbar & -G^{qq} & -\frac{1}{2}i\hbar - G^{qp} & 0 & 0 & 0
\end{array} \right)
\]
is invertible.  The Dirac bracket $\left\{ f, g \right\}_{\rm
Dirac}:=\left\{ f, g \right\} - \left\{ f, \phi_i \right\}
(\Delta^{-1})^{ij} \left\{ \phi_j, g \right\}$ for a second class
system of constraints can easily be computed for the remaining free
parameters $G^{qq}$, $G^{pp}$ and $G^{qp}$, recovering precisely the
algebra satisfied by the physical quantum variables (\ref{physqv}).
Thus, fixing the gauge freedom entirely, we recover the physical
Poisson algebra. In a general situation, where finding the explicit
form of observables is more difficult, this alternative method of
obtaining their Poisson algebra is easier to utilize.

\section{Truncations}
\label{s:Trunc}

Linear constraints show that consistency and completeness are
satisfied in our formulation of effective constraints. Locally, every
constraint can be linearized by a canonical transformation, but global
issues may be important especially in the quantum theory. Moreover,
moments transform in complicated ways under general canonical
transformations, mixing the orders of quantum variables. We will thus
discuss non-linear examples to show the practicality of our
procedures. Before doing so, we provide a more systematic analysis of
the treatment of infinitely many constraints as they arise on the
quantum phase space.

The above examples only considered quantum variables up to second
order. A reduction of this form is always necessary if one intends to
derive effective equations from a constrained system.  For practical
purposes, infinite dimensional systems have to be reduced to a certain
finite order of quantum variables so that one can actually retrieve
some information from the system. There are two possibilities to do
this: an approximate solution scheme order by order, or a sharp
truncation. It is then necessary to check whether the system of
constraints can still be formulated in a consistent way after such a
reduction has been carried out. A priori one cannot assume, for
instance, that a sharply truncated system of constraints has any
non-trivial solution at all. It may turn out that all degrees of
freedom are removed by the truncated constraints. Also it is not clear
how many (truncated) constraints have to be taken into account at a
certain order of the truncation.  In this section, we first consider a
linear example and show that it can consistently be truncated. We then
turn to the more elaborate and more physical example of the
parametrized free, non-relativistic particle. Here, sharp truncations
turn out to be inconsistent. While this makes sharp truncations
unreliable as a general tool, it is instructive to go through examples
where they are inconsistent. The following section will then be
devoted to consistent approximations without a sharp truncation.

\subsection{Truncated system of constraints for $\hat{C}=\hat{q}$}

The system as in Sec.~\ref{S:Cq} is governed by a constraint $C_{\rm
class}=q$ which on the quantum level entails the constraint operator
$\hat{C}=\hat{q}$. (We explicitly denote the classical constraint as
$C_{\rm class}$ because by our general rule we reserve the letter $C$
for the expectation value $\langle\hat{C}\rangle$.) This implies the
following constraints on the quantum phase space:
\ben C^{(n)}&=&\langle\hat{C}^n\rangle=C_{\rm
class}^n+\sum_{j=0}^{n-1}{n-1\choose j}C_{\rm
  class}^jG_{0,n-j}\quad,\quad
C_q^{(n)}=\langle\hat{q}\hat{C}^n \rangle=C^{(n+1)}\\
C_p^{(n)}&=&\langle\hat{p}\hat{C}^n \rangle=pC_{\rm
class}^n+p\sum_{j=0}^{n-1}{n\choose j}C_{\rm
class}^jG_{0,n-j}+\sum_{j=0}^{n-1}{n\choose j}\frac{C_{\rm
class}^j}{a_{n-j}}\left(G_{1,n+1}-
i\hbar\frac{(n-j)^2}{(n-j+1)}G_{0,n-j-1}\right)
\een where
$a_{n-j}$ are constant coefficients.  These are accompanied by similar
expressions of higher polynomial constraints, i.e.\ $C_{p^m}^{(n)}$
which are more lengthy in explicit form due to the reordering involved
in quantum variables.

The lowest power constraint yields
$C^{(1)}=C_{\rm class}\approx0$. Inserting this, the higher power
constraints reduce to
\ben C^{(n)}&\approx& G_{0,n}\quad,\quad C_q^{(n)}\approx G_{0,n+1}\\
C_p^{(n)}&\approx&
pG_{0,n}+\frac{1}{a_n}\left(G_{1,n}-i\hbar\frac{n^2}{(n+1)^2}G_{0,n-1}\right)
\,. \een
Now, a sharp truncation at
$N^{\mathrm{th}}$ order implies setting $G_{a,b}=0$ for all
$a+b>N$. As non-trivial constraints remain
\ben
C^{(n)}\vert_{N}&\approx& G_{0,n} \qquad \mbox{for all}\quad n\le N\\
C^{(n)}_p\vert_{N}&\approx& pG_{0,n}+
\frac{1}{a_n}\left(G_{1,n}-i\hbar\frac{n^2}{(n+1)^2}G_{0,n-1}\right)\qquad
\mbox{for
all} \quad n\le N-1\\
C^{(N)}_p\vert_{N}&\approx& pG_{0,N}+
\frac{1}{a_N}\left(-i\hbar\frac{N^2}{(N+1)^2}G_{0,N-1}\right)\qquad
\mbox{for}\quad n=N\ .
\een
Solving the quantum constraints $C^{(n)}\approx 0$ and inserting
the solutions into the constraints $C^{(n)}_p$, yields
\ben
 C^{(n)}_p\vert_{N}&\approx&\frac{1}{a_n}G_{1,n}\qquad 
\mbox{for all}\quad n\le N-1\\
C^{(n)}_p\vert_{N}&\approx& 0 \qquad \mbox{for all}\quad n\ge N.
\een

Thus we find that for the truncated system, $G_{0,n}$ are eliminated
through the constraints $C^{(n)}=0$, whereas the quantum variables
$G_{1,n}$ are eliminated through $C^{(n)}_p=0$. Higher polynomial
constraints fix all remaining moments except $G_{n,0}$: They can be
expanded as
\begin{eqnarray*}
 C_{p^k}^{(n)} &=& \sum_{i=0}^k\sum_{j=0}^n {k\choose i}{n\choose j}
 p^iC_{\rm class}^j \langle(\hat{p}-p)^{k-i}(\hat{q}-q)^{n-j}\rangle\\
&\approx& \sum_{i=0}^k {k\choose i} p^i
 \langle(\hat{p}-p)^{k-i}(\hat{q}-q)^n\rangle=
 \frac{G_{k,n}}{b_{k,n}}+\cdots
\end{eqnarray*}
with some coefficients $b_{k,n}$ and where moments of lower order in
$p$ are not written explicitly because they can be determined from
constraints of smaller $k$.  Due to the constraint $C^{(1)}=C_{\rm
class}\approx 0$, moreover, expectation values are restricted to the
classical constraint hypersurface. No further restrictions on these
degrees of freedom arise and also the gauge flows act in the proper
way. In particular, all remaining unconstrained $G_{n,0}$ become pure
gauge. (This again confirms considerations in Sec.~\ref{S:Number}
because the gauge flow of $C_{q^m}^{(n)}=C^{(n+m)}$ is sufficient to
remove all gauge without making use of $C_{p^m}^{(n)}$ with $m\not=0$,
where operators not commuting with the constraint would occur.) The
system can thus be truncated consistently.  For a truncation at
$N^{\mathrm{th}}$ order of a linear classical constraint, constraints
up to order $N$ have to be taken into account.

However, the linear case is quite special because we only had to
truncate the system of constraints, but not individual constraints:
any effective constraint contains quantum variables of only one fixed
order.  Referring back to section~\ref{S:Number}, when $\hat{C}$ is
linear, we can impose all of the constraints and remove all of the
gauge degrees of freedom in variables up to a given order without
invoking higher-order constraints. This is accomplished by treating
higher-order constraints as imposing conditions on higher-order
quantum variables (possibly in terms of the lower-order unconstrained
variables) and noting that using Eq.~(\ref{eq:F_gauge_flow}) there is
no need to refer to constraints of order above $F^{a_1, a_2,\ldots,
a_{2N}; b, 0}$ itself in order to demonstrate that it is a pure-gauge
variable. The gauge-invariant degrees of freedom that remain weakly
commute with \emph{all} constraints and not just the constraints up to
the order considered; see Eq.~(\ref{Cfg}).  As a result, in the linear
examples of Sec.~\ref{Examples}, higher order constraints do not
affect the reduction of the degrees of freedom for orders below and so
could be disregarded without making any approximations.  For
non-linear constraints, however, orders of moments mix and constraints
relevant at low orders can contain moments of higher order. It is then
more crucial to see how the higher moments could be disregarded
consistently, as we will do in what follows.

\subsection{Truncated system of constraints for the parametrized free non-relativistic 
  particle}
\label{S:FreeTrunc}

The motion of a free particle of mass $M$ in one dimension is
described on the phase space $(p,q)$. Through the introduction of an
arbitrary time parameter $t$, time can be turned into an additional
degree of freedom. The system is then formulated on the
$4$-dimensional phase space with coordinates $(t,p_t;q,p)$.  The
Hamiltonian constraint of the parametrized free non-relativistic
particle is given by
\be
C_{\rm class}=p_t+\frac{p^2}{2M}\ ,
\ee
which is constrained to vanish.

Promoting phase space variables to operators, Dirac constraint
quantization yields the quantum constraint
\be
\lb{Schroedinger}
\left(\hat{p}_t+\frac{\hat{p}^2}{2M}\right)\Psi=0\ .
\ee
In the Schr\"odinger representation, one arrives at an equation
that is formally equivalent to the time-dependent Schr\"odinger
equation\footnote{In contrast to the ordinary, time-dependent
Schr\"odinger equation, time is an operator in the equation obtained
here and not an external parameter. This implies that the Hamiltonian
which generates evolution in time, $\hat{\mathcal{H}}_{\rm phys}=
\frac{\hat{p}^2}{2M}$, has the same action on physical states as the
momentum operator canonically conjugate to time. In contrast to the
physical Hamiltonian, which is bounded below and positive
semidefinite, the spectrum of the time momentum $\hat{p}_t$ covers the
entire real line. On physical solutions, however, only positive
``frequencies'' contribute.}
\be
i\hbar\frac{\del\Psi(t,q)}{\del t}=\frac{\hbar^2}{2M}
\frac{\del^2\Psi(t,q)}{\del q^2}\ .
\ee
As is well known, solutions to this equation are given by
\be \label{packet}
\Psi(t,q)=\int\md k A(k)e^{\frac{\mathrm{i}}{\hbar}E(k)t+\mathrm{i}kq}
\ee
where $E(k)=\frac{\hbar^2 k^2}{2M}$.

For the quantum variables we use, as before, the notation
\be G^{a,b}_{c,d}=\langle(\hat p -p)^a (\hat q-q)^b (\hat p_t
-p_t)^c(\hat t -t)^d\rangle_{\rm Weyl}\ .  \ee
In their general form, the
set of constraints on the quantum phase space is given in the Appendix.

\subsubsection{Zeroth order truncation}
Truncation of the system at zeroth order, i.e. setting all quantum
variables to zero, yields $C^{(n)}\vert_{N=0}=C_{\rm class}^n$
together with
\[
 C^{(n)}_q\vert_{N=0}=qC_{\rm
class}^n+\frac{\mathrm{i}\hbar}{2}n\frac pmC_{\rm class}^{n-1}\quad,\quad
C^{(n)}_t\vert_{N=0}=tC_{\rm class}^n+
\frac{\mathrm{i}\hbar}{2}nC_{\rm class}^{n-1}
\]
as the required constraints.  This truncation is {\em not}
consistent. Inserting the condition $C_{\rm class}=0$ from the first
in the remaining constraints, especially in $C_t^{(1)}\vert_{N=0}=
tC_{\rm class}+\frac{1}{2}i\hbar$, results in
$\frac{\mathrm{i}\hbar}{2}=0$. The reason may seem clear: A truncation
at zeroth order can be understood as neglecting all quantum properties
of the system. But this is not possible for a free particle. There is
no solution in which e.g., both, spread in $p$ and $q$ would be
negligible throughout the particle's evolution. There is no
wave-packet which would remain tightly peaked throughout the evolution
and a description in terms of expectation values alone seems
insufficient in this case.

\subsubsection{Second order truncation}
\label{s:SecOrdInCons}

But even if one takes into account the second order quantum variables,
spreads and fluctuations, an inconsistent system results. The expanded
constraints can also be found in the appendix, which we now sharply
truncate at second order in moments.
From $C^{(n)}$ only three non-trivial constraints follow
\ben C^{(1)}&=&C_{\rm class}+\frac{1}{2M}G^{2,0}_{0,0}\\
C^{(2)}|_{N=2}&=& C_{\rm class}^2-\left(6C_{\rm class}-4p_t\right)
G^{2,0}_{0,0}+\frac{4p}{2M}G^{1,0}_{1,0}+G^{0,0}_{2,0}\\
C^{(3)}|_{N=2}&=& C_{\rm class}^3\ , \een upon inserting the
constraints successively. Thus for an $N=2$ order truncation, at
$n=3$, the classical constraint is recovered and must vanish for the
truncated system. Then, $C^{(1)}\approx0$ yields
$G^{2,0}_{0,0}\approx0$ which is too strong for a consistent reduction
since one expects the fluctuation $G^{pp}$ to be freely
specifiable. It has to remain a physical degree of freedom after
solving the constraints, for otherwise no general wave packet as in
(\ref{packet}) can be posed as an initial condition of the free
particle. In the sharp truncation, it turns out, there are too many
constraints which overdetermine the system. Especially the constraint
$C^{(3)}$, when truncated to second order moments, reduces to the
classical constraint $C_{\rm class}^3$, which then immediately implies
$G^{pp}=0$ due to $C^{(1)}$.

This observation points to a resolution of the inconsistency: While
$C^{(1)}$ is already of second order even without a truncation,
$C^{(3)}$ contains higher order moments. The truncation is then
inconsistent in that we are ignoring higher orders next to an
expression which we then constrain to be zero. For unconstrained
moments, this would be consistent; but it is not if some of the
moments are constrained to vanish. Thus, a more careful approximation
scheme must be used where we do not truncate sharply but ignore higher
moments only when they appear together with lower moments {\em not
constrained to vanish}. In such a scheme, as discussed in the
following section, $C^{(3)}$ would pose a constraint on the higher
moments in terms of $C_{\rm class}\approx -G^{pp}/2M$, but would not
require $C_{\rm class}$ or $G^{pp}$ to vanish.

\section{Consistent approximations}
\label{s:Approx}

Through the iteration described in Sec.~\ref{IterationConstraints},
the polynomial constraints of Sec.~\ref{S:Number} or the generating
function of Sec.~\ref{s:Generate} one arrives at an infinite number of
constraints imposed on an infinite number of quantum variables. The
linear systems have already demonstrated consistency and completeness
of the whole system, but for practical purposes the infinite number of
constraints and variables is to be reduced. We have seen in the
preceding section that sharp truncations are in general inconsistent
and that more careful approximation schemes are required. Depending on
the specific reduction, it is neither obvious that the effective
constraints are consistent in that they allow solutions for
expectation values and moments at all, nor is it guaranteed that the
constraints at hand do actually eliminate all unphysical degrees of
freedom. For each classical canonical pair which is removed by
imposing the constraints, all the corresponding moments as well as
cross-moments with the unconstrained canonical variables should be
removed. Classically, as well as in our quantum phase space
formulation, the elimination of unphysical degrees of freedom is a
twofold process: The constraints can either restrict unphysical
degrees of freedom to specific functions of the physical degrees of
freedom, or unphysical degrees of freedom can be turned into mere
gauge degrees of freedom under the transformations generated by the
constraints and then gauge fixed if desired.

In the following, we will first demonstrate by way of a non-trivial
example, rather than referring to linearization, that the constraints
as formulated in Sec.~\ref{IterationConstraints} are consistent,
before turning to the elimination of the unphysical degrees of
freedom.  Our specific example is again the parametrized free
non-relativistic particle, but the general considerations of
Sec.~\ref{s:ConsCons} hold for any parameterized non-relativistic
system.

We use the variables and constraints as they have been determined in
Sec.~\ref{S:FreeTrunc}.  This establishes a hierarchy of the
constraints, suggesting to solve $C^{(n)}$ first, then $C^{(n)}_q$,
$C^{(n)}_t$, $C^{(n)}_{p_t}$ and $C^{(n)}_{p}$, and the remaining
constraints (\ref{Con5})--(\ref{Con9}) first for $k=1$, then $k=2$
etc.  Note that for each $k$ in (\ref{Con5})--(\ref{Con9}) the $r=k$
term is the only contribution of a form not appearing at lower
orders. The terms occurring in the $r$-sum are linear combinations of
the constraints (\ref{Con5})--(\ref{Con9}) for $k^\prime< k$. Thus
apart from the $r=k$ term all other terms vanish if the lower $k$
constraints are satisfied.

It is important to notice that the structure of the constraints is
such that on the constraint hypersurface $C^{(n)}$, $C^{(n)}_{qp^k}$,
$C^{(n)}_{q}$, $C^{(n)}_{tp^k}$ and $C^{(n)}_{t}$ contain as lowest
order terms expectation values, whereas $C^{(n)}_{pp^k}$,
$C^{(n)}_{p}$, $C^{(n)}_{p_tp^k}$ and $C^{(n)}_{p_t}$ have second
order moments as lowest contribution. The highest order moments
occurring in $C^{(n)}$ are of order $2n$, $2n+1$ for $C^{(n)}_{q}$,
$C^{(n)}_{t}$, $C^{(n)}_{p}$ and $C^{(n)}_{p_t}$ and $2n+1+k$ in
$C^{(n)}_{qp^k}$, $C^{(n)}_{tp^k}$, $C^{(n)}_{pp^k}$ and
$C^{(n)}_{p_tp^k}$.

The structure of (\ref{Con5})--(\ref{Con9}) implies that the lowest
contributing order in the $j$- and $\ell$-sums (on the constraint
hypersurface) is $j+\ell+k\pm1$ and rises with $k$.  Consequently,
there exists a maximal $k$ up to which constraints have to be studied
if only moments up to a certain order are taken into account.  We
check the consistency of the constraints order by order in the
moments. This means that we first have to verify that one can actually
solve the constraints for the expectation values. This analysis will
then be displayed explicitly for second and third order moments.

\subsection{General procedure and moment expansion}
\label{s:ConsCons}

To verify consistency up to a certain order, one can exploit the fact
that up to a fixed order $N$ of the moments only a finite number of
constraints have to be taken into account. This can be seen from the
following argument: In the $j$-$\ell$-summation, the relevant moments
occur for $j+\ell\pm 1\le N$. From this condition, a number of pairs
$(j,\ell)$ result for which the sums occurring in
(\ref{Con1})--(\ref{Con9}) can be evaluated. There remain sums over
$m$ containing $p_t$, which should be eliminated if we choose $t$ as
internal time to make contact with the quantum theory of the
deparameterized system. (Our consistent approximation procedure,
however, is more general and does not require the choice of an
internal time.) We can achieve this by rewriting these as terms of the
form $n(n-1)\cdots(n-g)C_{\rm class}^{n-g-1}$ multiplied by powers of
$p$ and $2M$, where $g$ is an integer depending on the values of $j$
and $\ell$. (See the examples in
Eqs.~(\ref{Constraint1})--(\ref{Constraint7c}).) This is achieved by
eliminating $p_t$ via\footnote{In our example of the free particle, we
have $C_Q=p_t+p^2/2M+ G^{2,0}_{0,0}/2M$. If there is a potential,
there will be further classical terms as well as quantum variables
$G^{0,n}_{0,0}$.}  $C^{(1)}=C_Q\approx 0$ and illustrates the central
role played by the principal quantum constraint $C_Q$. For a fixed
order $N$ of moments, there is a factor of lowest and one of highest
power of $C_{\rm class}$. In $C^{(n)}$, e.g., the highest power is
given for $j=0$, $\ell=0$ (with $m=n$) and is simply $C_{\rm
class}^n$, whereas the lowest power is given for $\ell=0$, $j=N$ and
is given by $n(n-1)\cdots(n-(N-1))C_{\rm class}^{n-N}$.\footnote{This
term arises of course as well for $\ell=N,j=0$, $\ell=1,j=N-1$, etc.}

Since $C_{\rm class}\approx -G^{2,0}_{0,0}/2M$, powers of second order
moments ensue (or higher $q$-moments if there is a
potential). Together with powers of $\hbar$ in some of the terms, this
must be compared with the orders of higher moments in order to
approximate consistently. To formalize the required {\em moment
expansions}, one can replace each moment $G^{a,b}_{c,d}$ by
$\lambda^{a+b+c+d}G^{a,b}_{c,d}$ and expand in $\lambda$. This
automatically guarantees that higher order moments appear at higher
orders in the expansion, and that products of moments are of higher
order than the moments themselves. Moreover, in order to leave the
uncertainty relation unchanged, we have to replace $\hbar$ by
$\lambda^2\hbar$, which ensures that it is of higher order, too,
without performing a specific $\hbar$-expansion. After the
$\lambda$-expansion has been performed, $\lambda$ can be set equal to
one to reproduce the original terms. (Assumptions of orders of moments
behind this expansion scheme can easily be verified for Gaussian
coherent states of the harmonic oscillator, where a moment $G^{a,b}$
is of order at least $\hbar^{(a+b)/2}$.)

One can now rewrite the sum over $m$ for all those terms which produce
factors with powers of $C_{\rm class}$ down to the lowest power
occurring in front of the relevant moments. In $C^{(n)}$ this would
correspond to $C_{\rm class}^{n-N}$. One can therefore rewrite the
constraints in the form
\be  \label{constrexp}
C_{\rm class}^n Y_1+nC_{\rm class}^{n-1} Y_2+n(n-1)C_{\rm
class}^{n-2} Y_3+ \cdots +R\approx 0\ , 
\ee 
where $Y_i$ are functions linear in moments including those of order
smaller than $N$, and $R$ contains only moments which are of higher
order.  This allows one to successively solve the constraints for
$n=1$, $n=2$, etc. and discard all constraints arising for $n\ge N+1$,
$n>0$. In each case, one has to find the terms of lowest order in the
moment expansion, in combination with powers $C_{\rm class}^n$, to see
at which order a constraint becomes relevant.

It is crucial for this procedure to work that $C_{\rm class}^n$, which
arises in all constraints, can be eliminated at least for all
$n>n^\prime$ through terms of higher order moments using the principal
constraint $C_Q$. This key property can easily be seen to be realized
for any non-relativistic particle even in a potential, as long as
$p_t$ appears linearly. (For relativistic particles, additional
subtleties arise as discussed in a forthcoming paper.) While
(\ref{Con1})--(\ref{Con9}) change their form in such a case with a
different classical constraint, the procedure sketched here still
applies. Thus, it does not only refer to quadratic constraints but is
sufficiently general for non-relativistic quantum mechanics.

We will explicitly demonstrate the procedure for the free particle in
what follows. For that purpose, we rewrote the set of constraints in
the required form (\ref{constrexp}) for moments up to third order as
seen in App.~\ref{SetOfConstraints}.

\subsection{Consistency of constraints for expectation values}
\label{s:ExpValCons}

At zeroth order, we keep only expectation values. All moments are of
order $\mathcal{O}(\lambda^2)$ or higher. As only relevant constraints
we therefore find $C^{(n)}\approx 0$,
cf. App.~\ref{SetOfConstraints}. Keeping only zeroth order terms, this
reduces to $C^{(n)}=C^n_{\rm class}\approx 0$. This in turn
corresponds to the single constraint $C_{\rm class}\approx 0$ which
can be used to eliminate $p_t$ in terms of $p$. The system of
constraints is obviously consistent at zeroth order and no constraints
on variables associated with the pair $(q,p)$ result.

As explained above, the only constraint that restricts zeroth order
moments is $C^{(1)}=C_{\rm class}\approx 0$. This constraint allows us
to eliminate $p_t$. It generates a gauge flow on expectation values
given by \be \dot p=0\ , \qquad \dot p_t=0\ , \qquad \dot q=\frac pM\
, \qquad \dot t=1\ .  \ee The two observables of the system are
therefore
\begin{equation} \label{QP0}
{\cal P}^{(0)}=p\quad\mbox{and}\quad
{\cal Q}^{(0)}=q-t\frac pM \quad\mbox{with}\quad
\{{\cal Q}^{(0)},{\cal P}^{(0)}\}=1\ .
\end{equation}
These correspond to the two physical degrees of freedom corresponding
to expectation values of canonical variables. Among the four original
degrees of freedom of the system, $p_t$ is eliminated via the
constraint and $t$ is a pure gauge degree of freedom. There are no
further constraints to this order, which is thus consistent.

\subsection{Consistency of constraints up to second order moments}

At second order, we include second order moments and orders of $\hbar$
(recall that $\hbar$ is of order $\lambda^2$ in the moment expansion)
in addition to expectation values. Third order contributions are set
to zero.  We find that in addition to $C^{(1)}$, the new constraints
$C^{(1)}_q$, $C^{(1)}_t$, $C^{(1)}_{p_t}$ and $C^{(1)}_p$ arise. All
other constraints are of higher order: Second order moments enter in
these equations only through quadratic terms or with a factor of
$\hbar$, both of which are considered as higher order terms, cf.\
App.~\ref{SetOfConstraints}.  The only non-trivial constraints are
therefore
\bea
C^{(1)}&=&C_{\rm class}+\frac{1}{2M}G^{2,0}_{0,0}\approx 0\\
C^{(1)}_q&=&G^{0,1}_{1,0}+\frac pM\frac{\mathrm{i}\hbar}{2}+
\frac pM G^{1,1}_{0,0}\approx 0
\lb{SecOrdConstraints2}\\
C^{(1)}_t&=&\frac pM G^{1,0}_{0,1}+
G^{0,0}_{1,1}+\frac{\mathrm{i}\hbar}{2}\approx 0
\lb{SecOrdConstraints3}\\
C^{(1)}_{p_t}&=&G^{0,0}_{2,0}+\frac pM G^{1,0}_{1,0}\approx 0
\lb{SecOrdConstraints4}\\
C^{(1)}_p&=&G^{1,0}_{1,0}+\frac pM G^{2,0}_{0,0}\approx 0
\lb{SecOrdConstraints5}\ ,
\eea
where third order contributions have been set to zero. In accordance
with the order of expectation values, we use the first constraint to
eliminate $p_t=-p^2/2M-G^{2,0}_{0,0}/2M$ and solve for second
order moments
\bea
\lb{SecOrderConstraints}
G^{0,1}_{1,0}&=&- \frac pM\frac{\mathrm{i}\hbar}{2}-
\frac pM G^{1,1}_{0,0}\quad,\quad
\frac pM G^{1,0}_{0,1}=
-G^{0,0}_{1,1}-\frac{\mathrm{i}\hbar}{2}\\
G^{0,0}_{2,0}&=&-\frac pM G^{1,0}_{1,0}
\quad,\quad
G^{1,0}_{1,0}=-\frac pM G^{2,0}_{0,0}\,.
\nonumber
\eea
As constraints for $k>1$ contain second order moments only through
$C^n$, they are trivial as well. This follows from the first
constraint which sets $C^n\sim
\left(G^{2,0}_{0,0}\right)^n\sim\mathcal{O}(\lambda^{2n})$.  Thus, as
far as the second order moments are concerned, the system of
constraints is consistent: $G^{0,0}_{2,0}$, $G^{1,0}_{1,0}$,
$G_{0,1}^{1,0}$ and $G_{1,0}^{0,1}$ are fully determined while all
second order moments associated with the pair $(q,p)$ can be specified
freely. All remaining constraints then determine higher moments. This
is the same situation as experienced in the linear case as far as
solving the constraints for second order moments is concerned. The
inconsistency of Sec.~\ref{s:SecOrdInCons} is avoided because
$C^{(3)}$, which made $C_{\rm class}$ and thus $G^{2,0}_{0,0}$ vanish
in the sharp truncation, is now realized as a higher order constraint
in the moment expansion.

Gauge transformations are generated by $C^{(1)}$, $C^{(1)}_q$,
$C^{(1)}_t$, $C^{(1)}_{p_t}$ and $C^{(1)}_p$ where third order
contributions are set to zero as in (\ref{SecOrderConstraints}).  In
comparison to Sec.~\ref{s:ExpValCons} we have four additional
gauge transformations. Whereas ${\cal P}^{(2)}:={\cal P}^{(0)}$
remains gauge invariant under these transformations as well, this is
not the case for ${\cal Q}^{(0)}$. The latter has to be alleviated by
adding second order moments such that an observable
\be \label{Q2}
{\cal Q}^{(2)}={\cal Q}^{(0)}-\frac{1}{M}G^{1,0}_{0,1}
\ee
results satisfying $\{{\cal Q}^{(2)},{\cal P}^{(2)}\}=1$.

Calculating the transformations generated by the constraints on second
order moments shows that ${\cal G}^{pp}{}^{(2)}=G^{2,0}_{0,0}$ is an
observable, i.e. commutes with all five constraints on the
hypersurface defined by these constraints. The form of the gauge
orbits suggests to make the ansatz
\begin{eqnarray}
{\cal G}^{qp}{}^{(2)}&=&G^{1,1}_{0,0}+G^{0,0}_{1,1}-
\frac{t}{M}G^{2,0}_{0,0}+\frac{\mathrm{i}\hbar}{2}\label{Gqp2}\\
{\cal G}^{qq}{}^{(2)}&=&G^{0,2}_{0,0}-2\frac
pMG^{0,1}_{0,1}+\frac{p^2}{M^2}G^{0,0}_{0,2}
-\frac{2t}{M}\left(G^{1,1}_{0,0}+G_{1,1}^{0,0}+
\frac{\mathrm{i}\hbar}{2}\right)+\frac{t^2}{M^2}G^{2,0}_{0,0}
\end{eqnarray}
for the remaining two observables. They are invariant under gauge
transformations. The term $\frac{\mathrm{i}\hbar}{2}$ is included such
that the Poisson brackets between ${\cal G}^{qq}{}^{(2)}$ and the
remaining two quantum observables are of the required form.
They satisfy
\[
\{{\cal G}^{pp}{}^{(2)},{\cal G}^{qp}{}^{(2)}\}=-2{\cal
  G}^{pp}{}^{(2)}
\quad,\quad
\{{\cal G}^{pp}{}^{(2)},{\cal G}^{qq}{}^{(2)}\}=-4{\cal
  G}^{qp}{}^{(2)}
\quad,\quad
\{{\cal G}^{qp}{}^{(2)},{\cal G}^{qq}{}^{(2)}\}=-2{\cal G}^{qq}{}^{(2)}\,.
\]
Commutators between the variables ${\cal Q}^{(2)}$, ${\cal P}^{(2)}$
and the physical quantum variables ${\cal G}^{qq}{}^{(2)}$, ${\cal
G}^{pp}{}^{(2)}$ and ${\cal G}^{qp}{}^{(2)}$ vanish.

Thus we showed that four of the ten second order moments are
eliminated directly by the constraints. Three of the remaining second
order moments, $G^{0,0}_{1,1}$, $G^{0,0}_{0,2}$ and $G^{0,1}_{0,1}$,
are pure gauge degrees of freedom. Consequently three physical quantum
degrees of freedom remain at second order.  The observables can be
used to determine the general motion of the system in coordinate time:
From (\ref{QP0}) and (\ref{Q2}) together with
(\ref{SecOrderConstraints}) and (\ref{Gqp2}) we obtain
\begin{eqnarray}
 q(t)&=&{\cal Q}^{(2)}+\frac{t}{M}{\cal P}^{(2)}+ \frac{1}{M}G^p_t\approx {\cal
 Q}^{(2)}+\frac{t}{M}{\cal P}^{(2)}-\frac{1}{p}
 \left(G_{tp_t}+\frac{\mathrm{i}\hbar}{2}\right)\nonumber \\
 &=&  {\cal
 Q}^{(2)}+\frac{t}{M}{\cal P}^{(2)} - \frac{1}{{\cal P}^{(2)}} \left({\cal
 G}^{qp}{}^{(2)}+ \frac{t}{M} {\cal G}^{pp}{}^{(2)}-
 G^{qp}\right) \label{qoft}
\end{eqnarray}
for the relational dependence between $q$, $t$ and $G^{qp}$. 
Thus, the moments appear in the solutions for expectation values in
coordinate time which illustrates the relation between expectation
values and moments. At this stage, we still have to choose a gauge if
we want to relate the non-observables $q$, $t$ and $G^{qp}$ in this
equation to properties in a kinematical Hilbert space. A convenient
choice is to treat $(t,p_t)$ like a fully constrained pair as we have
analyzed it in the example of a linear constraint in
Sec.~\ref{Examples}. This suggests to fix the gauge by requiring that
$G_{tp_t}=-\frac{1}{2}i\hbar$ has no real part but only the imaginary
part for physical quantum variables to be real. Moreover, as in the
linear case we can gauge fix $G_{tt}=0$, such that the uncertainty
relation $G_{tt}G_{p_tp_t}-(G_{tp_t})^2\geq \hbar^2/4$ is saturated
independently of the behavior of the $(q,p)$-variables. (For
$G_{tt}\not=0$, it would depend on those variables via
$G_{p_tp_t}\approx p^2G^{pp}/M^2$ from (\ref{SecOrderConstraints}).)
Finally, this is the only gauge condition for $G_{tp_t}$ which works
for all values of ${\cal P}^{(2)}$, including ${\cal P}^{(2)}=0$ in
(\ref{qoft}).

In this gauge, we obtain
\begin{equation}
 q(t)= {\cal Q}^{(2)} +\frac{{\cal P}^{(2)}}{M}t \quad,\quad
 G^{qp}(t)= {\cal G}^{qp}{}^{(2)}+ \frac{{\cal G}^{pp}{}^{(2)}}{M}t
\end{equation}
in agreement with the solutions one would obtain for the
deparameterized free particle. In this case, there is no quantum
back-reaction of quantum variables affecting the motion of expectation
values because the particle is free. In the presence of a potential,
equations analogous to those derived here would exhibit those
effects. While it would in general be difficult to determine precise
observables in such a case, they can be computed perturbatively
starting from the observables found here for the free particle.

\subsection{Consistency of constraints up to third order moments}

Including third order terms in the analysis, solutions to the
constraints $C^{(1)}_q$, $C^{(1)}_t$, $C^{(1)}_{p_t}$ and $C^{(1)}_p$
become
\bea
G^{0,1}_{1,0}&=&-\frac pM\frac{\mathrm{i}\hbar}{2}-
\frac pM G^{1,1}_{0,0}-\frac{1}{2M}G^{2,1}_{0,0}
\lb{ThirdOrdConstraints2}\\
\frac pM G^{1,0}_{0,1}&=&
-G^{0,0}_{1,1}-\frac{\mathrm{i}\hbar}{2}-\frac{1}{2M}G^{2,0}_{0,1}
\lb{ThirdOrdConstraints3}\\
G^{0,0}_{2,0}&=&-\frac pM G^{1,0}_{1,0}-
\frac{1}{2M}G^{2,0}_{1,0}\lb{ThirdOrdConstraints4}\\
G^{1,0}_{1,0}&=&-\frac pM G^{2,0}_{0,0}-
\frac{1}{2M}G^{3,0}_{0,0}\lb{ThirdOrdConstraints5}\ .
\eea
As in the previous subsection, they will be used to determine second
order moments. The constraint $C^{(1)}$ contains no third order term
and thus remains unaltered. Third order moments are determined by
higher constraints $C^{(1)}_{qp}$, $C^{(1)}_{tp}$, $C^{(1)}_{p_tp}$,
$C^{(1)}_{p^2}$ and $C^{(2)}_q$, $C^{(2)}_t$, $C^{(2)}_{p_t}$. All
other constraints contain third order moments with a factor of $\hbar$
or of second or higher moments, both of which provides terms of higher
order. For instance, we may consider the constraints $C^{(1)}_{qp^2}$,
$C^{(1)}_{tp^2}$, cf. (\ref{Constraint6b}), (\ref{Constraint7b}). They
both contain third order moments with a factor of $C_{\rm class}$,
which, after solving $C^{(1)}$, becomes a term of fifth order. The
remaining second and third order terms occur with a factor of $\hbar$,
and are thus of fourth and fifth order. From this consideration of
orders in the moment expansion we conclude that $C^{(1)}_{qp^2}$ and
$C^{(1)}_{tp^2}$ do not constrain third order moments but become
relevant only at higher than third orders of the approximation scheme.

For $n=1$ the constraints that actually determine third order moments
are $C^{(1)}_{qp}$, $C^{(1)}_{tp}$, $C^{(1)}_{p_tp}$ and
$C^{(1)}_{p^2}$. On the constraint hypersurface, they imply
\ben
G^{1,1}_{1,0}&\approx&-
\frac{p}{M}G^{2,1}_{0,0}+
\frac{1}{2M}G^{2,0}_{0,0}\left(G^{1,1}_{0,0}-\mathrm{i}\hbar\right)\quad,\quad
G^{1,0}_{1,1}\approx\frac{1}{2M}G^{2,0}_{0,0}G^{1,0}_{0,1}-
\frac{p}{M}G^{2,0}_{0,1}\\
G^{1,0}_{2,0}&\approx&\frac{1}{2M}G^{2,0}_{0,0}\left(
\frac{1}{2M}G^{3,0}_{0,0}+\frac pM G^{2,0}_{0,0} \right)-
\frac{p}{M}G^{2,0}_{1,0}\quad,\quad
G^{2,0}_{1,0}\approx\frac{1}{2M}G^{2,0}_{0,0}G^{2,0}_{0,0}-
\frac{p}{M}G^{3,0}_{0,0}\ .
\een
Note that this holds on the constraint hypersurface defined by the
constraints $C^{(1)}$, $C^{(1)}_q$, $C^{(1)}_t$, $C^{(1)}_{p_t}$ and
$C^{(1)}_p$.  Dropping fourth and fifth order terms, we find the
simple relations
\[
G^{1,1}_{1,0}\approx-\frac{p}{M}G^{2,1}_{0,0}\quad,\quad
G^{1,0}_{1,1}\approx-\frac{p}{M}G^{2,0}_{0,1}\quad,\quad
G^{1,0}_{2,0}\approx-\frac{p}{M}G^{2,0}_{1,0}\quad,\quad
G^{2,0}_{1,0}\approx-\frac{p}{M}G^{3,0}_{0,0}\, .
\]
This happens in a consistent manner because unconstrained third order
moments appear on the right hand sides.  No condition for the
$(q,p)$-moments appearing here arises in this way, but the third order
moments $G_{1,1}^{1,0}$ and $G_{2,1}^{0,0}$ associated with $(t,p_t)$
remain unspecified at this stage. The constraints $C^{(2)}_q$,
$C^{(2)}_t$, $C^{(2)}_{p_t}$ arising for $n=2$ yield
\ben
G^{0,1}_{2,0}&\approx&\frac{p}{2M^2}G^{2,0}_{0,0}G^{1,1}_{0,0}\\
G_{2,1}^{0,0}&\approx&\frac1M\left(G^{2,0}_{0,0}\left(G^{0,0}_{1,1}+
\frac{1}{2M}G^{2,0}_{0,1}\right)+\frac{p^2}{M}G^{2,0}_{0,1}\right)\\
G_{3,0}^{0,0}&\approx&2\frac{p}{M}\left(-\frac{p^2}{2M^2}G^{3,0}_{0,0}+
\frac{1}{2M}G^{2,0}_{0,0}\left(\frac{1}{2M}G^{3,0}_{0,0}+
\frac{p}{2M}G^{2,0}_{0,0}\right)\right)\ ,
\een
which, after setting higher order terms to zero, sets
\[
G^{0,1}_{2,0}\approx 0\quad,\quad
G_{2,1}^{0,0}\approx\frac{p^2}{M^2}G^{2,0}_{0,1}\quad,\quad
G_{3,0}^{0,0}\approx -2\frac{p^3}{2M^3}G^{3,0}_{0,0}\, .
\]

The inclusion of third order terms and new constraints does not affect
${\cal P}^{(2)}$ and ${\cal Q}^{(2)}$. They remain constant under
gauge transformations. We therefore write
\be
{\cal P}^{(3)}:={\cal P}^{(0)}\quad\mbox{and}\quad{\cal Q}^{(3)}
:={\cal Q}^{(2)}\ .
\ee
Accordingly, their Poisson bracket is unaltered. The situation is
different for the second order quantum variables.  Only ${\cal
G}^{pp}{}^{(2)}$ remains invariant under the flow generated by third
order constraints.  Now that third order terms are included, ${\cal
G}^{qp}{}^{(2)}$ and ${\cal G}^{qq}{}^{(2)}$ are no longer
observables.  The former transforms under gauge transformations as
follows
\ben
\{{\cal G}^{qp}{}^{(2)},C^{(1)}_q\}&=&\frac{1}{2M}G^{2,1}_{0,0}\quad,\quad
\{{\cal G}^{qp}{}^{(2)},C^{(1)}_t\}=\frac{1}{2M}G^{2,0}_{0,1}\\
\{{\cal G}^{qp}{}^{(2)},C^{(1)}_{p_t}\}&=&\frac{1}{2M}G^{2,0}_{1,0}\quad,\quad
\{{\cal G}^{qp}{}^{(2)},C^{(1)}_p\}=\frac{1}{2M}G^{3,0}_{0,0}
\een
whereas Poisson brackets with $C^{(1)}_{qp}$, $C^{(1)}_{tp}$,
$C^{(1)}_{p_tp}$ and $C^{(1)}_{p^2}$ are of fourth order in the moment
expansion.  The terms on the right hand side can be eliminated through
the addition of a third order moment by
\be
{\cal G}^{qp}{}^{(3)}:={\cal G}^{qp}{}^{(2)}-\frac{1}{2M}G^{2,0}_{0,1}\ .
\ee
This has vanishing Poisson brackets with all constraints up to fourth
order terms. Moreover, it has vanishing Poisson bracket with ${\cal
P}^{(3)}$ as well as ${\cal Q}^{(3)}$. The Poisson bracket with ${\cal
G}^{pp}{}^{(3)}:= {\cal G}^{pp}{}^{(2)}$ remains unaltered, $\{{\cal
G}^{qp}{}^{(3)},{\cal G}^{pp}{}^{(3)}\}=2{\cal G}^{pp}{}^{(3)}$.

The transformations generated by the constraints on ${\cal
  G}^{qq}{}^{(2)}$ are of a more complicated form and we have not
  found a simple way of writing ${\cal G}^{qq}{}^{(3)}$ in explicit
  form. We conclude at this place because the applicability of
  effective constraints has been demonstrated. As already mentioned,
  the procedure also applies to interacting systems: We can solve the
  constraints in the same manner and using the same orders of
  constraints. The main consequence in the presence of a potential
  $V(q)$ is that additional $q$-moments appear as extra terms in
  solutions at certain orders, whose precise form depends on the
  potential. For a small potential, this can be dealt with by
  perturbation theory around the free solutions.


\section{Conclusions}

We have introduced an effective procedure to treat constrained
systems, which demonstrates how many of the technical and conceptual
problems arising otherwise in those cases can be avoided or
overcome. The procedure applies equally well to constraints with zero
in the discrete or continuous parts of their spectra and is, in fact,
independent of many representation properties. For each classical
constraint, infinitely many constraints are imposed on an
infinite-dimensional quantum phase space comprised of expectation
values and moments of states. This system is manageable when solved
order by order in the moments because this requires the consideration
of only finitely many constraints at each order. A formal definition
of this moment expansion has been given in Sec.~\ref{s:ConsCons}.

The principal constraint is simply the expectation value
$C_Q=\langle\hat{C}\rangle$ of a constraint operator, viewed as a
function of moments via the state used. Unless the constraint is
linear, there are quantum corrections depending on moments which can
be analyzed for physical implications. Moments are themselves subject
to further constraints, thus restricting the form of quantum
corrections in $C_Q$.

We have demonstrated that there is a consistent procedure in which an
expansion by moments can be defined, in analogy with an expansion by
moments in effective equations for unconstrained systems. This has
been shown to be applicable to any parameterized non-relativistic
system. We have also demonstrated the procedure with explicit
calculations in a simple example corresponding to the parameterized
free non-relativistic particle. In these cases, when faced with
infinitely many constraints we could explicitly choose an internal
time variable and eliminate all its associated moments to the orders
considered. Especially for the free particle, we were able to
determine observables invariant under the flows generated by the
constraints, and more generally observed how such equations encode
quantum back-reaction of moments on expectation values in an
interacting system.  These observables were subjected to reality
conditions to ensure that they correspond to expectation values and
moments computed in a state of the {\em physical Hilbert space}.
Especially physical Hilbert space issues appear much simpler in this
framework compared to a direct treatment, being imposed just by
reality conditions for functions rather than self-adjointness
conditions for operators. Nevertheless, crucial properties of the
physical Hilbert space are still recognizable despite of the fact that
we do not refer to a specific quantum representation. We also
emphasize that we choose an internal time after quantization, because
we do so when evaluating effective constraints obtained from
expectation values of operators. This is a new feature which may allow
new concepts of emergent times given by quantum variables even in
situations where no classical internal time would be available (see
e.g.\ \cite{Recollapse}).

In the examples, we have explicitly implemented the physical Hilbert
space by reality conditions on observables given by physical
expectation values and physical quantum variables. Observables thus
play important roles and techniques of
\cite{PartialCompleteObs,PartialCompleteObsII,DittrichThesis} might
prove useful in this context. Notice that we are referring to
observables of the quantum theory, although they formally appear as
observables in a classical-type theory of infinitely many constraints
for infinitely many variables. The fact that it often suffices to
compute these observables order by order in the moment expansion
greatly simplifies the computation of observables of the quantum
theory. Nevertheless, especially for gravitational systems of
sufficiently large complexity one does not even expect classical
observables to be computable in explicit form. Then, additional
expansions such as cosmological perturbations can be combined with the
moment expansion to make calculations feasible. This provides almost
all applications of interest. Moreover, if observables cannot be
determined completely, gauge fixing conditions can be used. As we
observed, depending on the specific gauge fixing some of the
kinematical quantum variables (before imposing constraints) can be
complex-valued while the final physical variables are required to be
real. Different gauge fixings imply different kinematical reality
conditions, which can be understood as different kinematical Hilbert
space structures all resulting in the same physical Hilbert space.

While we have discussed only the simplest examples, this led us to
introduce approximation schemes which are suitable more generally. In
more complicated systems such as quantum cosmology one may not be able
to find, e.g., explicit expressions for physical quantum variables as
complete observables. But for effective equations it is sufficient to
know the local behavior of gauge-invariant quantities, which can then
be connected to long-term trajectories obtained by solving effective
equations. A local treatment, on the other hand, allows one to
linearize gauge orbits, making it possible to determine observables.
Moreover, as always in the context of effective equations, simple
models can serve as a basis for perturbation theories of more
complicated systems.

A class of systems of particular interest is given by quantum
cosmology as an example for parameterized relativistic systems to be
discussed in a forthcoming paper. In such a case, the linear term
$p_t$ in the systems considered here would be replaced by a square
$p_t^2$. There is thus a sign ambiguity in $p_t$ which has some subtle
implications. Moreover, the principal quantum constraint $C_Q$ will
then acquire an additional moment $G_{p_tp_t}$ which may spoil the
suitability of $t$ as internal time in quantum theory provided that
the fluctuation $G_{p_tp_t}$ can become large enough for no real
solution for $p_t$ to exist. This demonstrates a further advantage of
the effective constraint formalism which we have not elaborated here:
the self-consistency of emergent time pictures can be analyzed
directly from the structure of equations.  Finally, if there are
several classical constraints, anomaly issues can be analyzed at the
effective level without many of the intricacies arising for constraint
operators. Also this will be discussed in more detail elsewhere
\cite{ScalarGaugeInv}.

To summarize, we have seen that the principal constraint $C_Q$ already
provides quantum corrections on the classical constrained
variables. The procedure thus promises a manageable route to derive
corrections from, e.g., quantum gravity in a way in which physical
reality conditions can be implemented. Since such conditions can be
imposed order by order in moments as well as other perturbations,
results can be arrived at much more easily compared to the computation
of full physical states in a Hilbert space. Nevertheless, all physical
requirements are implemented.

\section*{Acknowledgements}

We thank Alejandro Corichi for discussions. B.S.\ thanks
the Friedrich-Ebert-Stiftung for financial support.
Work of M.B.\ was supported in part by NSF grant PHY0653127.

\appendix
\section{System of constraints for the parametrized free particle}
\lb{SetOfConstraints}

General expression for the constraints are
\begin{eqnarray}
C^{(n)}&=&\Mmjl G_{j,0}^{\ell,0}\label{Con1}\\
C^{(n)}_q&=&\Mmjl\left(
  qG_{j,0}^{\ell,0}+G_{j,0}^{\ell,1}+\frac{\mathrm{i}\hbar}{2}\ell G_{j,0}^{\ell-1,0}\right)\nonumber\\
C^{(n)}_t&=&\Mmjl\left(tG_{j,0}^{\ell,0}+G_{j,1}^{\ell,0}+\frac{\mathrm{i}\hbar}{2}jG_{j-1,0}^{\ell,0}\right)\nonumber\\
C^{(n)}_{p_t}&=&\Mmjl\left(p_tG_{j,0}^{\ell,0}+G_{j+1,0}^{\ell,0}\right)\label{Con4}\\
C^{(n)}_{p^k}&=&\Mmjlr G_{j,0}^{\ell+r,0}\label{Con5}\\
C^{(n)}_{tp^k}&=&\Mmjlr\nonumber\\
& &\times\left(tG_{j,0}^{\ell+r,0}+G_{j,1}^{\ell+r,0}+\frac{\mathrm{i}\hbar}{2}jG_{j-1,0}^{\ell+r,0}\right)\label{Con7}\\
C^{(n)}_{qp^k}&=&\Mmjlr\nonumber\\
& &\times\left(qG_{j,0}^{\ell+r,0}+G_{j,0}^{\ell+r,1}+\frac{\mathrm{i}\hbar}{2}(\ell+r)G_{j,0}^{\ell+r-1,0}\right)\label{Con8}\\
C^{(n)}_{p_tp^k}&=&\Mmjlr\nonumber\\
& &\times\left(p_tG_{j,0}^{\ell+r,0}+G_{j+1,0}^{\ell+r,0}\right)\label{Con9}\ .
\end{eqnarray}
In addition to those written explicitly here, there are those
involving higher polynomials also in $q$, $t$ and $p_t$. The first two
types of those constraints are more lengthy due to reorderings in the
quantum variables. The constraints listed suffice for considerations
in this paper.

In a moment expansion, the leading terms of these constraints are
\small{
\bea
C^{(n)}&=&C_{\rm class}^n+
nC_{\rm class}^{n-1}\frac{1}{2M}G^{2,0}_{0,0}\nonumber\\
& &+n(n-1)C_{\rm class}^{n-2}\left[\frac{p^2}{2M^2}G^{2,0}_{0,0}+
\frac pM G^{1,0}_{1,0}+\frac 12
G^{0,0}_{2,0}+\frac{1}{2M}G^{2,0}_{1,0}+
\frac{p}{2M^2}G^{3,0}_{0,0}+\frac{1}{8M^2}G^{4,0}_{0,0}\right]\nonumber\\
& &+
n(n-1)(n-2)C_{\rm class}^{n-3}\left[\frac{p^2}{2M^2}G^{2,0}_{1,0}+\frac{p}{2M}G^{1,0}_{2,0}+\frac
  16 G^{0,0}_{3,0}+\frac{p^3}{6M^3}G^{3,0}_{0,0}+X_1\right]\nonumber\\
& &+
n(n-1)(n-2)(n-3)R_1=0
\lb{Constraint1}\\
C^{(n)}_q&=&qC^{(n)}+
nC_{\rm class}^{n-1}\left[\frac pM\frac{\mathrm{i}\hbar}{2}+\frac
pMG^{1,1}_{0,0}+G^{0,1}_{1,0}+\frac{1}{2M}G^{2,1}_{0,0}\right]\nonumber\\
& &+
n(n-1)C_{\rm class}^{n-2}\left[\frac{\mathrm{i}\hbar}{2}\frac 1M
\left(G^{1,0}_{1,0}+\frac{3p}{2M}G^{2,0}_{0,0}\right)+
\frac{p^2}{2M^2}G^{2,1}_{0,0}+\frac{p}{M}G^{1,1}_{1,0}+
\frac 12G^{0,1}_{2,0}+\frac{\mathrm{i}\hbar}{2}
\frac{1}{2M^2}G^{3,0}_{0,0}+X_2\right]\nonumber\\
& &+
n(n-1)(n-2)C_{\rm class}^{n-3}\left[\frac{\mathrm{i}\hbar}{2}
\left(\frac{p^2}{M^2}G^{1,0}_{1,0}+\frac{p^3}{2M^3}G^{2,0}_{0,0}+\frac{p}{2M}
G^{0,0}_{2,0}+\frac{3p}{2M^2}G^{2,0}_{1,0}+\frac{p^2}{M^3}G^{3,0}_{0,0}+
\frac{1}{2M}G^{1,0}_{2,0}\right)+X_3\right]\nonumber\\
& &+
n(n-1)(n-2)(n-3)C_{\rm class}^{n-4}\left[\frac{\mathrm{i}\hbar}{2}
\left(\frac{p^4}{6M^4}G^{3,0}_{0,0}+\frac{p^3}{2M^3}G^{2,0}_{1,0}+
\frac{p^2}{2M^2}G^{1,0}_{2,0}+\frac{p}{6M}G^{0,0}_{3,0}\right)+
X_4\right]\nonumber\\
& &+
n(n-1)(n-2)(n-3)(n-4)R_2=0
\lb{Constraint2}\\
C^{(n)}_t&=&tC^{(n)}+
nC_{\rm class}^{n-1}\left[\frac{\mathrm{i}\hbar}{2}+\frac
  pMG^{1,0}_{0,1}+G^{0,0}_{1,1}+\frac{1}{2M}G^{2,0}_{0,1}\right]\nonumber\\
& &+
n(n-1)C_{\rm class}^{n-2}\left[\frac{\mathrm{i}\hbar}{2}
\frac{1}{2M}G^{2,0}_{0,0}+\frac{p^2}{2M^2}G^{2,0}_{0,1}+
\frac{p}{M}G^{1,0}_{1,1}+\frac 12 G^{0,0}_{2,1}+X_5\right]\nonumber\\
& &+
n(n-1)(n-2)C_{\rm class}^{n-3}\left[\frac{\mathrm{i}\hbar}{2}
\left(\frac{p}{M}G^{1,0}_{1,0}+\frac 12 G^{0,0}_{2,0}+
\frac{p^2}{2M^2}G^{2,0}_{0,0}+\frac{p}{2M^2}G^{3,0}_{0,0}+
\frac{1}{2M}G^{2,0}_{1,0}\right)+X_6\right]\nonumber\\
& &+
n(n-1)(n-2)(n-3)C_{\rm class}^{n-4}\left[\frac{\mathrm{i}\hbar}{2}
\left(\frac{p^3}{6M^3}G^{3,0}_{0,0}+\frac{p^2}{2M^2}G^{2,0}_{1,0}+
\frac{p}{2M}G^{1,0}_{2,0}+\frac 16 G^{0,0}_{3,0}\right)+X_7\right]\nonumber\\
& &+n(n-1)(n-2)(n-3)(n-4)R_3=0
\lb{Constraint3}\\
C^{(n)}_{p_t}&=&p_tC^{(n)}+
nC_{\rm class}^{n-1}\left[\frac
  pMG^{1,0}_{1,0}+G^{0,0}_{2,0}+\frac{1}{2M}G^{2,0}_{1,0}\right]\nonumber\\
& &+
n(n-1)C_{\rm class}^{n-2}\left[\frac{p^2}{2M^2}G^{2,0}_{1,0}+
\frac{p}{M}G^{1,0}_{2,0}+\frac 12 G^{0,0}_{3,0}+X_8\right]\nonumber\\
& &+n(n-1)(n-2)R_4=0
\lb{Constraint4}\\
C^{(n)}_{p}&=&pC^{(n)}+
nC_{\rm class}^{n-1}\left[G^{1,0}_{1,0}+\frac{1}{2M}G^{3,0}_{0,0}+\frac
  pM G^{2,0}_{0,0}\right]\nonumber\\
& &+
n(n-1)C_{\rm class}^{n-2}\left[\frac{p^2}{2M^2}G^{3,0}_{0,0}+
\frac{p}{M}G^{2,0}_{1,0}+\frac 12 G^{1,0}_{2,0}+X_9\right]\nonumber\\
& &+
n(n-1)(n-2)R_5=0
\lb{Constraint5a}\\
C^{(n)}_{p^2}&=&2pC^{(n)}_p-pC^{(n)}+
C_{\rm class}^nG^{2,0}_{0,0}+
nC_{\rm class}^{n-1}\left[\frac{p}{M}G^{3,0}_{0,0}+G^{2,0}_{1,0}+
\frac{1}{2M}G^{4,0}_{0,0}\right]+
n(n-1)R_6=0
\lb{Constraint5b}\\
C^{(n)}_{tp}&=&tC^{(n)}_p+pC^{(n)}_t+
C^nG^{1,0}_{0,1}+
nC_{\rm class}^{n-1}\left[\frac{p}{M}G^{2,0}_{0,1}+
\frac{1}{2M}G^{3,0}_{0,1}+G^{1,0}_{1,1}\right]\nonumber\\
& &+
n(n-1)C_{\rm class}^{n-2}\left[\frac{\mathrm{i}\hbar}{2}\left(\frac pM
    G^{2,0}_{0,0}+G^{1,0}_{1,0}+\frac{1}{2M}G^{3,0}_{0,0}\right)+
X_{10}\right]\nonumber\\
& &+
n(n-1)(n-2)C_{\rm class}^{n-3}\left[\frac{\mathrm{i}\hbar}{2}
\left(\frac{p^2}{2M^2}G^{3,0}_{0,0}+\frac
    12 G^{1,0}_{2,0}+\frac{p}{M}G^{2,0}_{1,0}\right)+X_{11}\right]\nonumber\\
& &+
n(n-1)(n-2)(n-3)R_7=0
\lb{Constraint6a}\\
C^{(n)}_{qp}&=&qC^{(n)}_p+pC^{(n)}_q+
C_{\rm class}^n\left[G^{1,1}_{0,0}+\frac{\mathrm{i}\hbar}{2}\right]\nonumber\\
& &+
nC_{\rm class}^{n-1}\left[3\frac{\mathrm{i}\hbar}{2}\frac{1}{2M}
G^{2,0}_{0,0}+\frac{p}{M}G^{2,1}_{0,0}+G^{1,1}_{1,0}+\frac{1}{2M}
G^{3,1}_{0,0}\right]\nonumber\\
&
&+n(n-1)C_{\rm class}^{n-2}\left[\frac{\mathrm{i}\hbar}{2}
\left(\frac{3p^2}{2M^2}G^{2,0}_{0,0}+\frac{2p}{M}G^{1,0}_{1,0}+\frac
    12
    G^{0,0}_{2,0}+\frac{2p}{M^2}G^{3,0}_{0,0}+\frac{3}{2M}G^{2,0}_{1,0}\right)+X_{12}\right]\nonumber\\
&
&+n(n-1)(n-2)C_{\rm class}^{n-3}\left[\frac{\mathrm{i}\hbar}{2}
\left(\frac{3p^2}{2M^2}G^{2,0}_{1,0}+\frac{2p^3}{3M^3}G^{3,0}_{0,0}+\frac
    pM G^{1,0}_{2,0}+\frac 16
    G^{0,0}_{3,0}\right)+X_{13}\right]\nonumber\\
& &+n(n-1)(n-2)(n-3)R_8=0
\lb{Constraint7a}\\
C^{(n)}_{p_tp}&=&p_tC^{(n)}_p+pC^{(n)}_{p_t}+
C_{\rm class}^nG^{1,0}_{1,0}+
nC_{\rm class}^{n-1}\left[\frac pM G^{2,0}_{1,0}+G^{1,0}_{2,0}+
\frac{1}{2M}G^{3,0}_{1,0}\right]+
n(n-1)R_9=0
\lb{Constraint8a}\\
C^{(n)}_{p^3}&=&3pC^{(n)}_{p^2}-3p^2C^{(n)}_{p}+p^3C^{(n)}+
C_{\rm class}^nG^{3,0}_{0,0}+
nC_{\rm class}^{n-1}X_{14}+
n(n-1)R_{10}=0
\lb{Constraint5c}\\
C^{(n)}_{tp^2}&=&tC^{(n)}_{p^2}-p^2C^{(n)}_{t}+2pC^{(n)}_{tp}-2ptC^{(n)}_{p}\nonumber\\
& &+
C_{\rm class}^nG^{2,0}_{0,1}+
nC_{\rm class}^{n-1}\left[\frac{\mathrm{i}\hbar}{2}G^{2,0}_{0,0}+
X_{15}\right]\nonumber\\
& &+
n(n-1)C_{\rm class}^{n-2}\left[\frac{\mathrm{i}\hbar}{2}\left(\frac pM
    G^{3,0}_{0,0}+G^{2,0}_{1,0}\right)+X_{16}\right]\nonumber\\
& &+
n(n-1)(n-2)R_{11}=0
\lb{Constraint6b}\\
C^{(n)}_{qp^2}&=&qC^{(n)}_{p^2}-p^2C^{(n)}_{q}+2pC^{(n)}_{qp}-2pqC^{(n)}_{p}\nonumber\\
& &+C_{\rm class}^nG^{2,1}_{0,0}+
nC_{\rm class}^{n-1}\left[\frac{\mathrm{i}\hbar}{2}\left(3\frac
    pMG^{2,0}_{0,0}+2G^{1,0}_{1,0}+4\frac{1}{2M}G^{3,0}_{0,0}\right)+
X_{17}\right]\nonumber\\
& &+
n(n-1)C_{\rm class}^{n-2}\left[\frac{\mathrm{i}\hbar}{2}
\left(\frac{2p^2}{M^2}G^{3,0}_{0,0}+3\frac
    pM G^{2,0}_{1,0}+G^{1,0}_{2,0}\right)+X_{18}\right]\nonumber\\
& &+n(n-1)(n-2)R_{12}=0
\lb{Constraint7b}\\
C^{(n)}_{p_tp^2}&=&p_tC^{(n)}_{p^2}-p^2C^{(n)}_{p_t}+2pC^{(n)}_{p_tp}-2pp_tC^{(n)}_{p}
\nonumber\\
& &+C_{\rm class}^nG^{2,0}_{1,0}+
nC_{\rm class}^{n-1}X_{19}+
n(n-1)R_{13}=0
\lb{Constraint8b}\\
C^{(n)}_{tp^3}&=&tC^{(n)}_{p^3}+p^3C^{(n)}_{t}-3p^2C^{(n)}_{tp}+3^2ptC^{(n)}_{p}+3pC^{(n)}_{tp^2}-3ptC^{(n)}_{p^2}\nonumber\\
& &+
C_{\rm class}^nG^{3,0}_{0,1}+
nC_{\rm class}^{n-1}\left[\frac{\mathrm{i}\hbar}{2}G^{3,0}_{0,0}+X_{20}\right]+
n(n-1)R_{14}=0
\lb{Constraint6c}\\
C^{(n)}_{qp^3}&=&qC^{(n)}_{p^3}+p^3C^{(n)}_{q}-3p^2C^{(n)}_{qp}+3^2pqC^{(n)}_{p}+3pC^{(n)}_{qp^2}-3pqC^{(n)}_{p^2}\nonumber\\
& &+
C_{\rm class}^n\left[G^{3,1}_{0,0}+3\frac{\mathrm{i}\hbar}{2}G^{2,0}_{0,0}\right]\nonumber\\
& &+
nC_{\rm class}^{n-1}\left[\frac{\mathrm{i}\hbar}{2}\left(4\frac pM G^{3,0}_{0,0}+3G^{2,0}_{1,0}\right)+X_{21}\right]+
n(n-1)R_{15}=0
\lb{Constraint7c}\ ,
\end{eqnarray}}
where $X_i$ and $R_i$ are linear functions of higher, i.e. at least fourth, order moments.



\end{document}